\documentclass[preprint]{aastex}
\usepackage[latin1]{inputenc}
\usepackage{graphicx}
\graphicspath{ {./} }
\usepackage{amsmath}

\slugcomment{}

\shorttitle{}
\shortauthors{Guennou et al.}

\begin{document}

\title{Relative Abundance Measurements in Plumes and Interplumes\\}

\author{C. Guennou\altaffilmark{1,}\altaffilmark{2}, M. Hahn\altaffilmark{1}, and D.W. Savin\altaffilmark{1}}

\affil{\altaffilmark{1}Columbia Astrophysics Laboratory, Columbia University, MC 5247, 550 West 120th Street, \\New York, NY 10027}
\affil{\altaffilmark{2}Present address: Instituto de Astrofisica de Canarias,  C/ Vía Láctea, s/n, 38205, La Laguna, Tenerife, Spain and Departamento de Astrofísica, Universidad de La Laguna, E-38206 La Laguna, Tenerife, Spain}

\email{cguennou@iac.es}

\begin{abstract}
      
We present measurements of relative elemental abundances in plumes and interplumes. Plumes are bright, narrow structures in coronal holes that extend along open magnetic field lines far out into the corona. Previous work has found that in some coronal structures the abundances of elements with a low first ionization potential (FIP) $< 10$~eV are enhanced relative to their photospheric abundances. This coronal-to-photospheric abundance ratio, commonly called the FIP bias, is typically 1 for element with a high-FIP ($> 10$~eV). We have used EIS spectroscopic observations made on 2007 March 13 and 14 over an $\approx 24$ hour period to characterize abundance variations in plumes and interplumes. To assess their elemental composition, we have used a differential emission measure (DEM) analysis, which accounts for the thermal structure of the observed plasma. We have used lines from ions of iron, silicon, and sulfur. From these we have estimated the ratio of the iron and silicon FIP bias relative to that for sulfur. From the results, we have created FIP-bias-ratio maps. We find that the FIP-bias ratio is sometimes higher in plumes than in interplumes and that this enhancement can be time dependent. These results may help to identify whether plumes or interplumes contribute to the fast solar wind observed in situ and may also provides constraints on the formation and heating mechanisms of plumes.
        
\end{abstract}

\keywords{Sun: corona - Sun: UV radiation - Sun: elemental abundances - solar wind - plumes }

\section{Introduction}
\label{sec:intro}

Polar plumes are bright, narrow structures rooted in coronal holes. They extend along open magnetic field lines, expanding with altitude as the magnetic field diverges. In white light, plumes have been observed as radial structures~\citep{vandehulst1950, saito1965} that reach altitudes greater than $\approx 30 R_{\odot}$~\citep{deforest2001}. Plumes are generally denser and cooler than their surroundings and are likely heated at the base~\citep{wang1994}. An extensive review of plume plasma properties can be found in~\citet{wilhelm2011}. Despite many studies carried out in white light and extreme ultraviolet (EUV), there are still many open questions about plumes, such as their formation mechanism and their possible contribution to the fast solar wind. 

Because of projection effects, the three-dimensional nature of plumes is still controversial. The classical view considers plumes to be quasi-cylindrical structures formed by a small emerging magnetic bipole reconnecting with surrounding open magnetic field lines. Such plumes are known as ``beam'' plumes. A second type of plume, called a ``curtain-plume'', has been proposed by~\citet{gabriel2003}. These plumes correspond to a chance alignment along the line of sight (LOS) of a series of the roughly vertical, faint outflows along the boundaries of the magnetic network on the solar surface. Since these two plume populations present almost the same plasma properties~\citep{gabriel2009}, it is difficult to observationally distinguish the two types. 

It is now accepted that coronal holes are the origin of the fast solar wind~\citep{krieger1973}. However, it is unknown whether plumes or interplumes are the main contributor of material to the fast solar wind. Some recent results have found that the outflow velocity is smaller in plumes than in the surrounding interplume gas~\citep[][]{teriaca2003, raouafi2007}. Since the solar wind speed at low heights is higher in interplumes than in plumes, these authors suggest that interplume regions are the preferred channel for the acceleration of the material into the fast solar wind. Conversely, other studies measured higher outflow velocities in plumes than in interplumes and concluded that plumes are the main contributor to the fast solar wind,~\citep[][]{gabriel2003, gabriel2005}. Theoretical models of plumes extending into the solar wind have been developed by~\citet{wang1994} and \citet{habbal1995} and more recently by~\citet{pinto2009} and~\citet{velli2011}. However, neither theory nor observations have been able to conclusively determine the main contributor to the fast solar wind.      

One way to determine if plumes or interplumes feed the fast solar wind is by searching for corresponding material in the in-situ measurements. Coronal abundances have been found to vary significantly among some coronal structures. Since the composition of the solar corona becomes fixed at low heights, elemental abundances measurements at these heights can thus be potentially used to identify the sources of the fast solar wind.   

These coronal abundance variations are correlated with the first ionization potential (FIP) specific to each element. The abundances of low-FIP elements (FIP $< 10$ eV) appear to be enhanced relative to their photospheric abundances in some coronal structures. On the other hand, elements having a high-FIP value (FIP $> 10$ eV) exhibit coronal abundances close to photospheric. For an exhaustive review of the abundance measurements in the solar corona see~\citet{raymond1999},~\citet{feldman2005}, and ~\citet{feldman2007}. It is convenient to report this so-called FIP effect in terms of a FIP bias, which is defined as
\begin{equation}
f_{\rm X} = \frac{A_{\rm X}^{\rm c}}{A_{\rm X}^{\rm p}},
\end{equation}
where $A_{\rm X}^{\rm c}$ and $A_{\rm X}^{\rm p}$ are the coronal and photospheric abundances of element X, respectively.   
      
It is still unclear whether or not plumes are subject to a FIP effect. Measurements by~\citet{wilhelm1998} and~\citet{young1999} with the Coronal Diagnostic Spectrometer (CDS) on board the \textit{Solar and Heliospheric Observatory (SOHO)} concluded that there is a FIP effect in plumes. A more recent study by~\citet{curdt2008} using measurements from the \textit{SOHO}/Solar Ultraviolet Measurements of Emitted Radiation (SUMER) spectrometer found a small enhancement in plumes. Conversely, using data from \textit{SOHO}/CDS, \citet{delzanna2003} found no significant FIP effect in plumes. A time dependence of the FIP bias in plumes could possibly explain the discrepancies among these studies. Such behavior has been seen in measurements in active regions where the FIP bias gradually increased over time~\citep{feldman2003}.     

Here, we investigate the elemental composition of both plumes and interplumes and their evolution over an $\approx 24$ hour period. Our measurements are based on a differential emission measure (DEM) analysis~\citep{feldman2003, feldman2009}. The DEM provides a measure of the amount of LOS emitting material as a function of electron temperature. Our DEM analysis was carried out using many emission lines from the low-FIP elements iron (FIP $\approx 7.90$~eV) and silicon (FIP $\approx 8.15$~eV). Relative abundances measurements are derived using the intermediate-FIP element sulfur (FIP $\approx 10.36$~eV). 

The rest of this paper is organized as follows. In Section~\ref{sec:data}, we describe the observations and the data processing. The DEM analysis is reviewed in detail in Section~\ref{sec:met}. The results are described in Section~\ref{sec:fip_maps}. Systematic uncertainties are reviewed in Section~\ref{res_accuracy}. The results are discussed in Section~\ref{sec:conclusions}. Finally, we summarize in Section~\ref{sec:summary}.  

\section{Data Analysis}
\label{sec:data}

\subsection{Plume and interplume observations}
\label{subsec:observations}
\begin{figure*}
\begin{center}
\epsscale{0.5}
\plotone{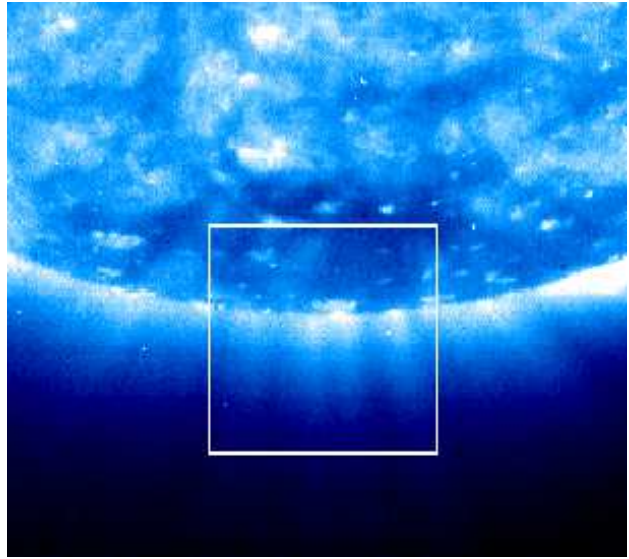}
\end{center}
\caption{The box outlines the raster range covered by the EIS 1$''$ slit observations on 2007 March 13 at 13:00 UT. The solar image is from the \textit{SOHO}/EIT 171~\AA\ observation closest in time to the EIS raster and consists mainly of emission from Fe X and Fe XI. \label{fig:fov_eis}}
\end{figure*}

We used off-limb observations from the Extreme Ultraviolet Imaging Spectrometer~\citep[EIS;][]{culhane2007} on \textit{Hinode}~\citep{kosugi2007}. The data were collected with the 1$''$ slit. We used 6 observations covering the full wavelength range and spanning an $\approx 24$ hour period. The first observation was performed on 2007 March 13 starting at 12:00~UT and ending at 15:39. The slit was rastered from $X=130.57''$ to $-124.13''$, respectively, and centered vertically at $Y=-972.43''$, with an exposure time of 50~s for each of the 256 pointings. Figure~\ref{fig:fov_eis} displays the positions of the slit, overlaid on the closest 171~\AA\ image from the \textit{SOHO}/Extreme Ultraviolet Imaging Telescope~\citep[EIT;][]{boudine1995}. We also used four smaller rasters made on this same day, starting respectively at 19:05, 20:07, 21:08 and 22:10. For these data, the 1$''$ slit was centered vertically at $Y=-972.42''$ and scanned from $X=30.80''$ to $-28.01''$ with a 60~s exposure time for each of the 60 pointings. The last observation was performed on 2007 March 14 from 11:05 to 14:44, with the same properties as the first observation. These off-limb rasters are ideal for our study because they present a well developed polar coronal hole and contain insignificant contamination from quiet Sun plasma (see Section~\ref{subsub:silicon_dem} for more details). 

The data were processed using standard routines available in the Interactive Date Language (IDL) \textit{Solar Software} (SSW) package. CCD dark current, spikes, and warm pixels were removed and the intensity was calibrated with the \texttt{eis\_prep.pro} routine. Next, missing pixels were interpolated, following the implementation of~\citet{young2010} and using the \texttt{eis\_replace\_missing.pro} procedure. Systematic drifts in the wavelength scale were then taken into account using \texttt{eis\_shift\_spec.pro}, based on the empirical model of~\citet{kamio2010}. The four smaller rasters on 2007 March 13 were coaligned and averaged together in order to improve their statistical accuracy. Finally, each of the three resulting data sets was summed in both the horizontal and vertical directions into spatial bins of 4 pixels by 4 pixels. 

\begin{figure*}
\begin{center}
\includegraphics[scale = 0.9, trim=450 150 450 90]{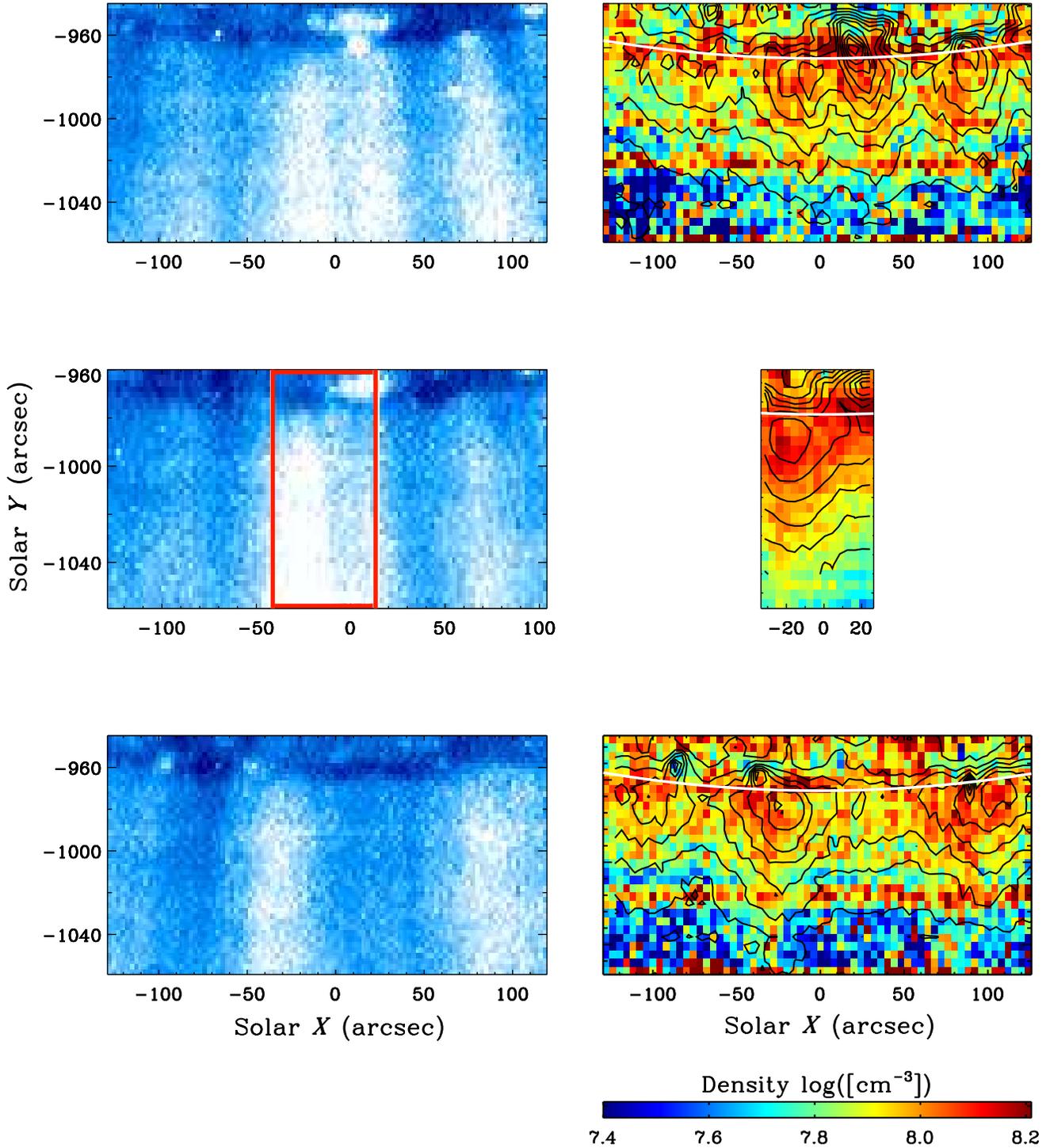}
\end{center}
\caption{\textit{Left column}: Maps of the measured intensity for each of the three data sets as seen by \textit{SOHO}/EIT in the 171~\AA\ channel. The red box on the middle pannel outlines the position of the EIS raster for this data set. \textit{Right column}: Maps of the measured density on a logarithmic scale for each of the three EIS data sets. The black contours correspond to the intensity measured using the Fe~X~184.54~\AA\ line. The solar limb is highlighted by the white line.\label{fig:density}}
\end{figure*}

The fields of view seen by EIT in the 171 \AA\ bandpass corresponding to our three data sets are shown in the left column of Figure~\ref{fig:density}. In the first observation, two different plumes are clearly visible in the center of the raster, another one on the right hand side, and a suggestion of a fourth one is visible on the left. One plume seems to come from a bright point, whereas the origin of the others is difficult to determine. The field of view is greatly reduced in the second data set, but one plume still falls within the raster. In the last observation, only three plumes are observed; the one coming from the bright point has completely disappeared. 

All of the observations used the full spectral range of EIS. This, combined with long exposure times, provided usable spectral lines covering a wide range of electron temperatures, between $\log T_{\rm e} = 5.7$ to $6.3$. Here and throughout, temperatures are given in units of Kelvin. We used 4 lines from \ion{Si}{7}, 2 from \ion{Si}{10}, and 15 from Fe charge states spanning \uppercase\expandafter{\romannumeral 8}--\uppercase\expandafter{\romannumeral 15}. These lines are mostly unblended (see Table~\ref{tab:eis_lines}). Measurements of relative abundances were derived using these 21 lines combined with the \ion{S}{8}~198.55 \AA\ line.

\subsection{Line fitting}
\label{subsec:lines}

We fit Gaussian profiles to each spectral line in order to derive intensities. For this, we used the \texttt{eis\_auto\_fit.pro} procedure, described in detail by~\citet{young2013}. Most of the lines chosen are unblended; in this case a single Gaussian fit was applied to the data. The initial EIS intensity error bars are derived from the IDL routine \texttt{eis\_prep.pro}, which include photon counting statistics and error estimates for the dark current. \citet{hahn2012} have found that this approach underestimates the uncertainties and so we adopted their Monte-Carlo procedure to derive the uncertainties for each of the Gaussian parameters used in the analysis here. Note that here and throughout, all uncertainties are given at an estimated $1\sigma$ statistical accuracy.

The few blended lines in our data set were easily dealt with. The \ion{Si}{7}~278.45~\AA\ line is blended with the \ion{Mg}{7}~278.39~\AA\ line. The lines were separated using a 2-Gaussian fit that forced them to have the same width~\citep{landi2009}. This assumption is reasonable because both the corresponding thermal and non-thermal widths of each ion are expected to be similar. The \ion{Fe}{12}~195.12~\AA\ is self-blended with the \ion{Fe}{12}~192.18~\AA\ line, leading to a distortion of the profile from a single Gaussian. For this line, we follow the procedure described by~\citet{young2009} and use a double Gaussian fit in order to derive the \ion{Fe}{12}~195.12~\AA\ intensity. The \ion{S}{8}~198.55~\AA\ line is blended with the \ion{Fe}{11}~198.54~\AA\ line. However, the \ion{Fe}{11} contribution can be estimated through the theoretical \ion{Fe}{11}~198.54~\AA/189.12~\AA\ branching ratio, which is $\approx$ 0.8~\citep{landi2009}. The \ion{Fe}{11}~189.12~\AA\ is unblended and relatively isolated from other lines, so the intensity can be accurately determined using a single Gaussian fit. We then estimate and subtract the \ion{Fe}{11}~198.54~\AA\ contribution to find the \ion{S}{8}~198.55~\AA\ intensity. Fortunately, the \ion{Fe}{11}~198.54~\AA\ intensity is very weak, with a ratio of \ion{Fe}{11}/\ion{S}{8} intensities which never exceeds 15\% for off-limb observations. Taking into account the estimated 10\% accuracy of the branching ratio, the resulting uncertainty in the final \ion{S}{8}~198.55~\AA\ intensity does not exceed 2\%.

We also took into account instrumental scattered light in our analysis. This light is superimposed on the true coronal line emission and may lead to an overestimation of the intensity, especially at high heights above the limb. To remove the stray light from our data, we subtracted a stray light line profile. The upper limit of the off-limb scattered light is estimated to be 2\% of the averaged on-disk intensities~\citep[][]{ugarte2010, hahn2012}. Thus, we first fit the lines with a single Gaussian for on-disk data with $R \leqslant 0.95\ R_{\odot}$, as measured from the Sun center. Then, we subtract from each off-limb spatial bin a Gaussian stray light line profile using the same centroid position and width and 2\% of the averaged on-disk intensity. The stray light in our analysed data does not exceed 2.5\% of the total measured for any given line intensity.

\subsection{Density diagnostic}
\label{subsec:density}

Some of the lines used in this study are density sensitive, such as the \ion{Fe}{9}~189.94~\AA\ and \ion{Fe}{12}~195.12~\AA\ lines. The intensity for these transitions depend on both temperature and density. This is a potential source of systematic uncertainty in the DEM inversion~\citep[][]{craig1976, brown1991, judge1997, guennou2012a, guennou2012b}. To mitigate this problem, we have determined the density using the intensity ratio of the \ion{Fe}{9}~188.50~\AA\ and 189.94~\AA\ lines~\citep[][]{young2009b}. For this, the plasma is assumed to be isothermal, at a temperature about $\log T_{\rm e} \approx 5.9$, corresponding to the peak of the contribution function of the two lines. We used the atomic data from CHIANTI version 7.1~\citep{dere1997, landi2013}. The right column of Figure~\ref{fig:density} shows the density measured for our data sets as a function of the solar-$X$ and $Y$ coordinates. The black contours correspond to the intensity observed simultaneously in the \ion{Fe}{10}~184.54 \AA\ line. The inferred density is $1.3\pm 0.2 \times 10^{8}$ cm$^{-3}$ in plumes and $7.9 \pm 0.8 \times 10^{7}$ cm$^{-3}$ in interplumes at $R = 1.02\ R_{\odot}$. This agrees with typical values previously measured in plumes and interplumes~\citep[][]{wilhelm2011}.  
       
\section{Measuring FIP-bias ratios}
\label{sec:met}

A DEM analysis, by accounting for the temperature dependence of the observed intensities, is one way to measure spectroscopically the elemental composition of an emitting plasma~\citep{feldman2003}. For the work here, we first derived the DEM using the Fe lines. The DEM was reconstructed using the regularization method of~\citet{hannah2012}. Taking this DEM we predicted the Si line intensities and found good agreement with those observed. This implies that both ions have a similar FIP bias. Hence, we combined the Fe and Si data to compute the DEM for each spatial bin of our three data sets. To maximize the reliability of the DEM inversion, we imposed various criteria pertaining to the number, intensity, temperature coverage, and uncertainty of the spectral lines used (see Section~\ref{subsub:reliability_dem}). Based on this Fe \& Si DEM, we then computed the predicted intensities of the moderate-FIP S~VIII~198.55~\AA\ line. Finally, by comparing both the predicted and observed S~VIII intensities, we determined the ratio for the FIP bias of the low-FIP elements Fe and Si relative to that for S.

\subsection{Differential Emission Measure (DEM)}
\label{subsub:reliability_dem}

In an optically thin plasma, the intensity of a given spectral line corresponding to the transition from level $j$ to level $i$ for element X is given by 

\begin{equation}
\label{eq:int_ne}
I_{ji} = \frac{A_{\rm X} }{4\pi} \int G_{ji}(T_{\rm e}, n_{\rm e}) n_{\rm e}^2 dl, 
\end{equation}
where $dl$ is the differential along the LOS; $T_{\rm e}$ and $n_{\rm e}$ are the electron temperature and density, respectively; $A_{\rm X}$ is the abundance of X relative to hydrogen; and $G_{ji}$ is the contribution function, which is defined as 

\begin{equation}
\label{eq:contribution_fct}
G_{ji}(T_{\rm e}, n_{\rm e}) = A_{ji}\frac{n_j({\rm X}^{+m})}{n({\rm X}^{+m})}\frac{n({\rm X}^{+m})}{n({\rm X})}\frac{n(H)}{n_{\rm e}}\frac{1}{n_{\rm e}}.
\end{equation}
Here, $A_{ji}$ is the radiative transition rate, $n_j({\rm X}^{+m})$ represents the number density of ion ${\rm X}^{+m}$ in level $j$, $n({\rm X}^{+m})$ is the number density of ion ${\rm X}^{+m}$, $n({\rm X})$ is the number density of the element X, and $n(H)/n_{\rm e}$ is the abundance of hydrogen relative to that of electrons. 

Following~\citet{pottasch1963, pottasch1964}, Equation~(\ref{eq:int_ne}) can be reformulated as
\begin{equation}
\label{eq:dem}
I_{ji} = \frac{A_{\rm X}}{4\pi} \int_{0}^{+\infty} G_{ji}(T_{\rm e}, n_{\rm e}) \xi(T_{\rm e}) d\log T_{\rm e}, 
\end{equation}
where the DEM $\xi$ provides a measure of the amount of emitting plasma as a function of the electron temperature. The DEM is defined as $\xi(T_{\rm e}) =n_{\rm e}^2 dl/d\log T_{\rm e}$~\citep[][]{craig1976}.
 
Reliably inferring the DEM from a series of observations has proved to be a challenge, due to the inverse nature of the problem~\citep[][]{craig1976, brown1991, testa2012, guennou2013}. To ensure an accurate solution, many lines having low uncertainties and spanning a broad temperature range are needed. In order to automate the DEM inversion process for the large number of spatial bins in our three data sets, we have imposed several criteria to reject spatial bins for which the DEM is poorly constrained. We required at least 7 different lines to perform the DEM inversion for each spatial bin. In order to ensure adequate temperature coverage, we also require that there be at least 5 different charge states in the available lines used to compute the DEM. Since the FIP bias is deduced using the \ion{S}{8}~198.55~\AA\ line, which forms at about $\log T_{\rm e} = 5.8$, we required at least 3 different charge state spanning the temperature range $\log T_{\rm e} = [5.7,6.0]$. Additionally, for each spatial bin, lines are rejected if they are too weak, which we define as $I_{ji} <$ 10~ergs~cm$^{-2}$~s$^{-1}$~sr$^{-1}$. Lines were also considered too noisy and rejected if the signal-to-noise ratio was less than 3, i.e., $I_{ji} < 3 \sigma$, where $\sigma$ is the uncertainty of $I_{ji}$ (see Section~\ref{subsec:lines}). All together these conditions ensure that the analysis uses only spatial bins for which the DEM is well constrained.  
 
\subsection{The iron DEM}
\label{subsub:iron_dem}

The first step in our approach is to derive the DEM in the coronal hole using only the iron lines. The coronal DEM $\xi_{\rm c}$ is related to the observed Fe intensities through    
\begin{equation}
\label{eq:int_fe_obs}
I^{\rm Fe}_{\rm obs} =\it \frac{A_{\rm{Fe}}^{\rm c} }{ \rm 4\pi} \it \int_{0}^{+\infty} G_{ji}(T_{\rm e}, n_{\rm e}) \xi_{\rm c}(T_{\rm e}) d\log T_{\rm e}.
\end{equation}
Since the coronal abundance of low-FIP elements is unknown \textit{a priori}, we recast Equation~(\ref{eq:int_fe_obs}) in terms of the photospheric abundances which are believed to be known. Thus, the observed Fe intensities are
\begin{equation}
\label{eq:int_fe_dem}
I^{\rm Fe}_{\rm obs} =\it \frac{A_{\rm{Fe}}^{\rm p}}{ \rm 4\pi} \it \int_{0}^{+\infty} G_{ji}(T_{\rm e}, n_{\rm e}) \xi_{\rm p}(T_{\rm e}) d\log T_{\rm e},
\end{equation}
where we use the relationship $A_{\rm Fe}^{\rm c}=f_{\rm Fe}A_{\rm Fe}^{\rm p}$ and $\xi_{\rm p}$ is the coronal DEM derived using photospheric abundances. The DEM $\xi_{\rm p}$ is related to the coronal DEM $\xi_{\rm c}$ by
\begin{equation}
\label{eq:dem_obs}
\xi_{\rm p}(T_{\rm e}) = f_{\rm Fe} \it\ \xi_{\rm c}(T_{\rm e}).
\end{equation}
Using the approach of~\citet{hannah2012} we can invert Equation~(\ref{eq:int_fe_dem}) to find the inferred $\xi_{\rm p}$. Figure~\ref{fig:dem_ex} displays a typical example of such a DEM $\xi_{\rm p}$ and will be discussed in more detail in Section~\ref{subsub:silicon_dem}.

\subsection{The Fe-to-Si FIP-bias ratio}
\label{subsub:fe_si_ratio}

The next step is to compute the FIP-bias ratio $f_{\rm Fe}/f_{\rm Si}$. Since Si and Fe are both low-FIP elements, with similar FIP values, they are expected to have a similar FIP bias. Numerous observational studies have demonstrated this in measurements of many different coronal structures~\citep[][]{feldman2003, feldman2007, brooks2011}. In order to estimate the $f_{\rm Fe}/f_{\rm Si}$ ratio, we compare the predicted and observed intensities of the 6 Si lines used in this analysis. 

The predicted Si intensities are computed using the photospheric silicon abundance. As before the photospheric and coronal abundances are related by the silicon FIP bias $f_{\rm Si}$ as $A_{\rm{Si}}^{\rm c} = f_{\rm Si}A_{\rm{Si}}^{\rm p}$. Using the inferred DEM $\xi_{\rm p}$ derived from the Fe lines by inverting Equation~(\ref{eq:int_fe_dem}), the predicted Si intensities can be written as
\begin{equation}
\label{eq:int_si_pred}
I^{\rm Si}_{\rm pred} = \it f_{\rm Fe} \frac{A_{\rm{Si}}^{\rm p} }{\rm 4\pi} \it \int_{0}^{+\infty} G_{ji}(T_{\rm e}, n_{\rm e}) \xi_{\rm c}(T_{\rm e}) d\log T_{\rm e}, 
\end{equation}
where $\xi_{\rm p}$ has been recast using Equation~(\ref{eq:dem_obs}). The observed Si intensities are given by
\begin{equation}
\label{eq:int_si_obs}
I^{\rm Si}_{\rm obs} =\it f_{\rm Si} \frac{A_{\rm{Si}}^{\rm p} }{ \rm 4\pi} \it \int_{0}^{+\infty} G_{ji}(T_{\rm e}, n_{\rm e}) \xi_{\rm c}(T_{\rm e}) d\log T_{\rm e}.
\end{equation}
Therefore, the ratio of the predicted to observed Si intensity reduces to
\begin{equation}
\label{eq:ratio}
\frac{I^{\rm Si}_{\rm pred}}{I^{\rm Si}_{\rm obs}} = \frac{f_{\rm Fe}}{f_{\rm Si}}.
\end{equation}
If both Fe and Si have the same FIP bias, then the observed and computed Si intensities should match.  

The question now arises of which compilation of photospheric abundances to use for our analysis. Several sets are available. In order to determine which set of the reported photospheric abundances is consistent with our observations, we tested the effect on our measurements using the recommended values of~\citet{grevesse2007} and of \citet{asplund2009}. The silicon abundance measured in these two studies is identical whereas that for iron differs, having $(A_{\rm{Fe}})^{\rm Aspl}/(A_{\rm{Fe}})^{\rm Grev} = 1.122$. We find that the abundances of~\citet{asplund2009} give a FIP-bias ratio $f_{\rm Fe}/f_{\rm Si}$ of close to unity. This is as expected, and so we use the \citet{asplund2009} photospheric abundances for the rest of the analysis. 

\begin{figure*}
\begin{center}
\includegraphics[scale = 0.9, trim=450 150 700 90]{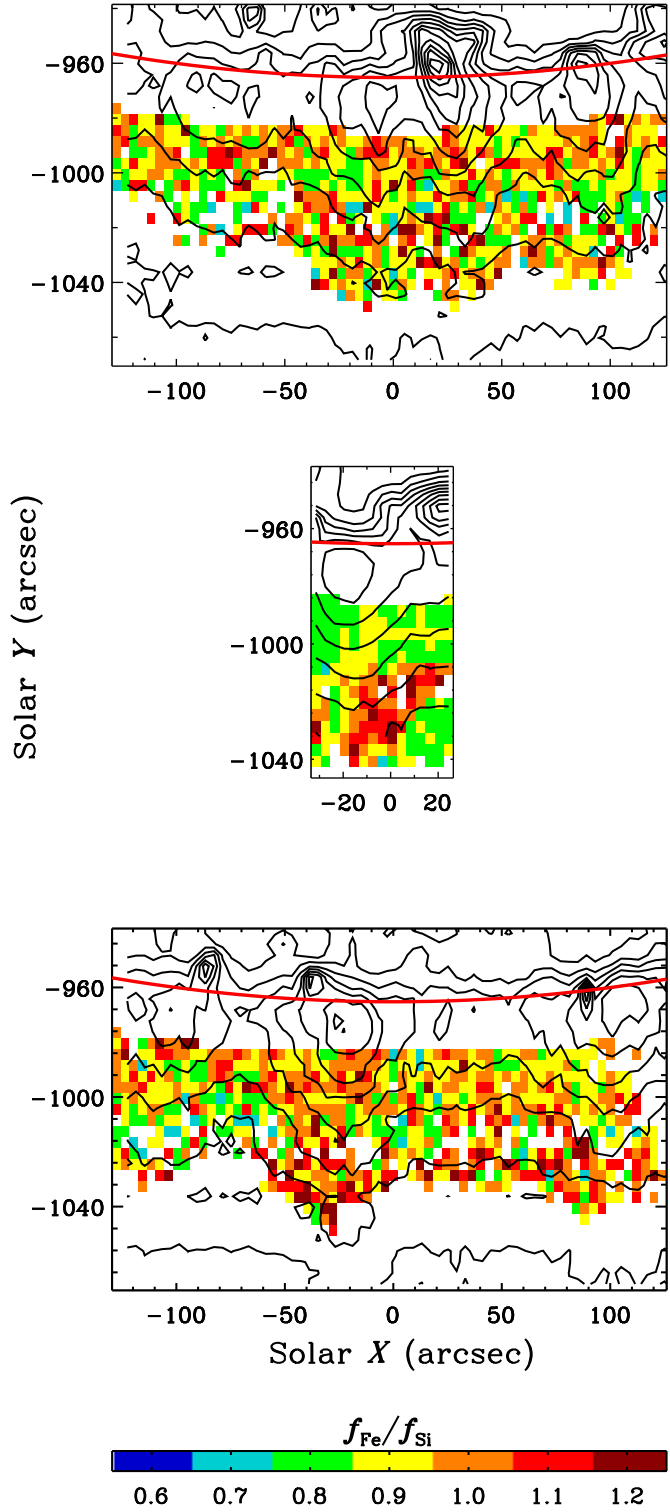}
\end{center}
\caption{Maps of the FIP-bias ratio $f_{\rm Fe}/f_{\rm Si}$ for each data set. The black contours are as in Figure~\ref{fig:density}. The red line displays the solar limb. The FIP-bias ratio is concentrated between 0.9 and 1.1, with mean values and $1\sigma$-standard deviations of $0.97 \pm 0.12$, $0.96 \pm 0.09$, and $0.96 \pm 0.10$, going from the top map to the bottom. \label{fig:si_maps}}
\end{figure*}

\begin{figure*}
\begin{center}
\includegraphics[scale = 0.9, trim=450 200 700 90]{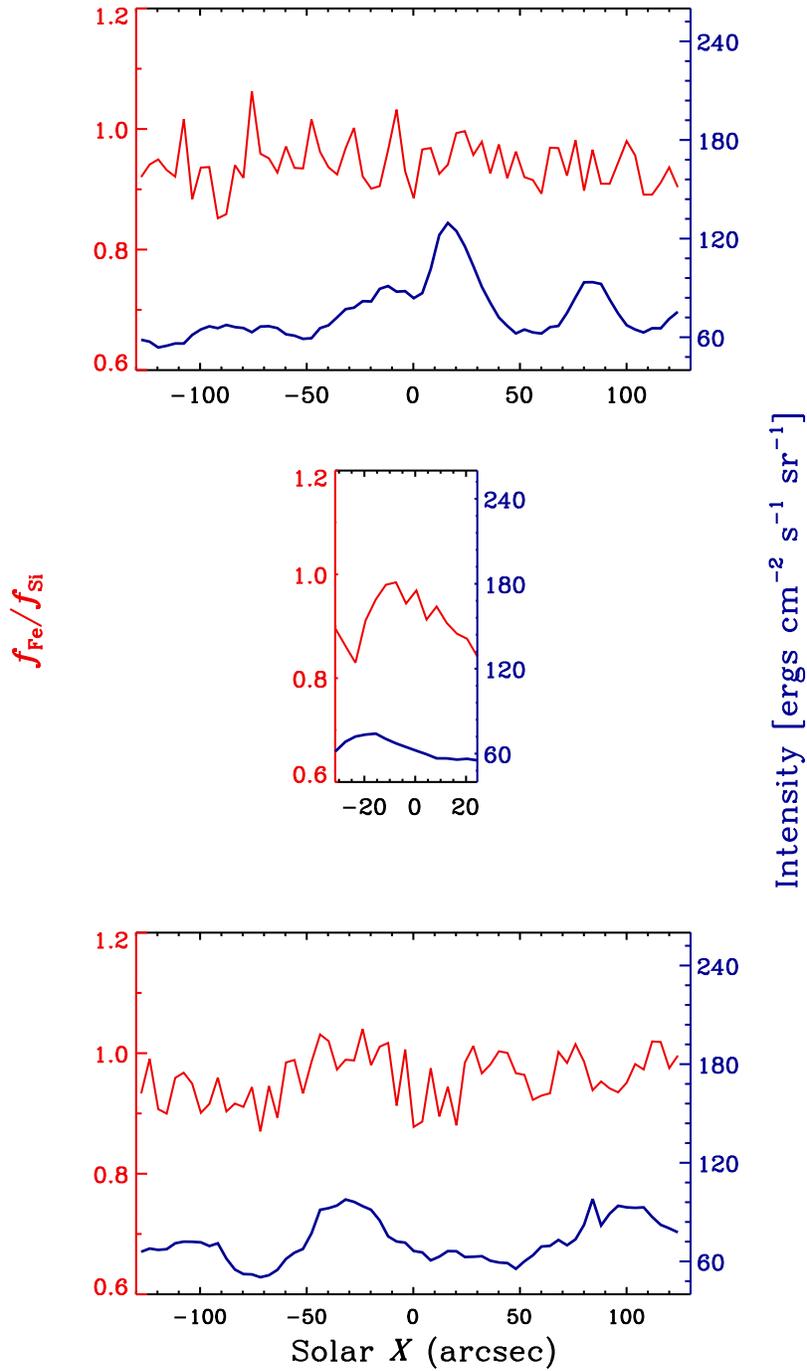}
\end{center}
\caption{FIP-bias ratio profiles $f_{\rm Fe}/f_{\rm Si}$ for each observation (red lines and left axes). Each profile has been computed by averaging the spatial bins of the maps in Figure~\ref{fig:si_maps} along the solar-$Y$ axis. In order to simultaneously present the plume positions, the observed intensities in the Fe X 184.54~\AA\ line are also displayed on the same plot (blue lines and right axes). These intensity profiles have been obtained by averaging 7 spatial bins from $Y\approx-1002''$ to $Y\approx-1030''$ in order to enhance the signal and increase the visibility of the features. \label{fig:profiles_fe}}
\end{figure*}

Figure~\ref{fig:si_maps} shows the variation of the $f_{\rm Fe}/f_{\rm Si}$ ratio for each of our three data sets. The white areas correspond to rejected spatial bins, following the criteria we described in Section~\ref{subsub:reliability_dem}. We also rejected data for which the radius is $R < 1.02\ R_{\odot}$. At such heights, spicules are present, which can lead to a multithermal distribution and therefore confuse the analysis. The corresponding profiles as a function of the solar-$X$ coordinate are displayed in Figure~\ref{fig:profiles_fe}, along with the intensity profiles observed using the Fe X 184.54 \AA\ line to identify the position of each plumes. For each data set, we find a relatively homogeneous $f_{\rm Fe}/f_{\rm Si}$ ratio of between 0.9 and 1.1. There is no clear correlation between the measured $f_{\rm Fe}/f_{\rm Si}$ ratio and the \ion{Fe}{10} intensity. It is worth noting that these maps are sparser than those presented later in this work. This is because we have used only iron lines to compute the DEM, and so fewer spatial bins were consistent with the criteria that we applied to ensure the reliability of the DEM inversion.

\subsection{The Fe \& Si DEM}
\label{subsub:silicon_dem}

\begin{figure*}
\begin{center}
\includegraphics[angle=270, scale = 0.7]{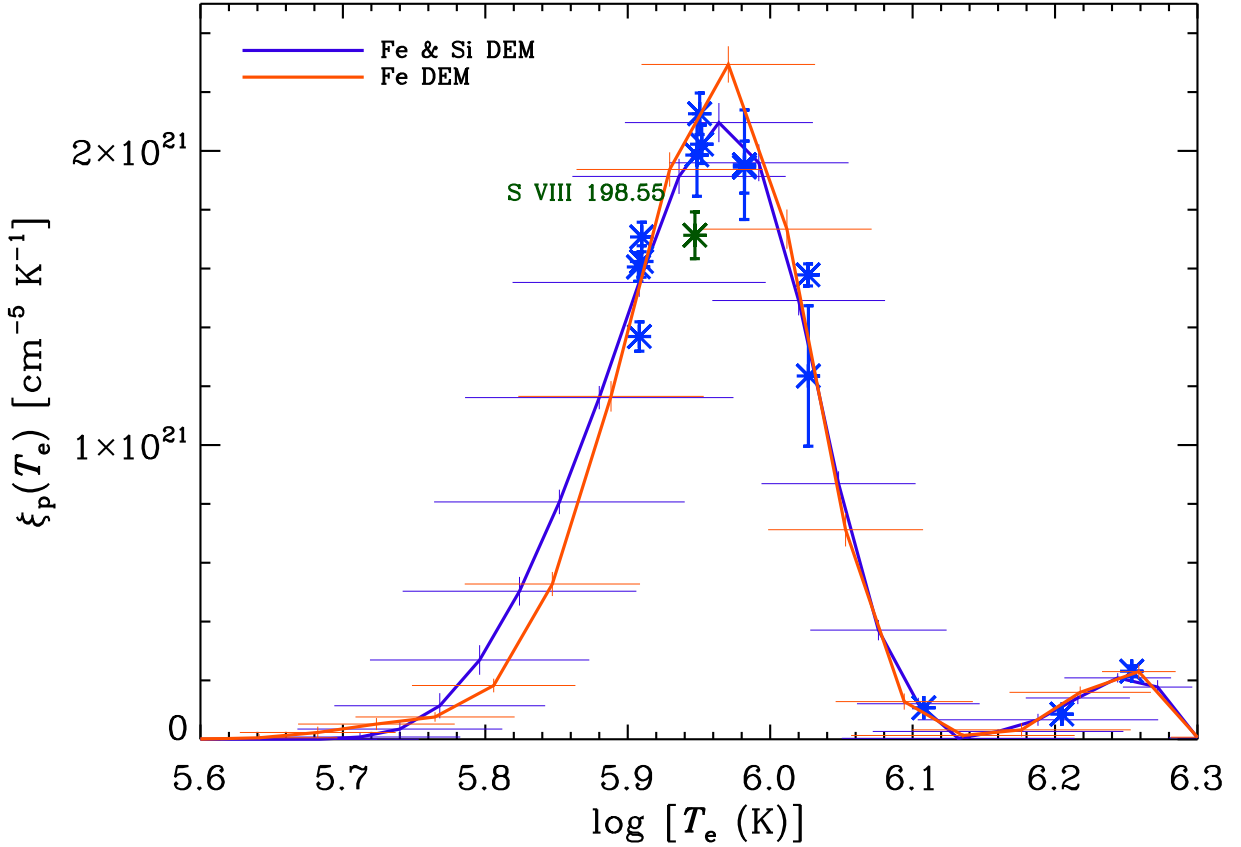}
\end{center}
\caption{Example of a typical $\xi_{\rm p}$ DEMs computed for one spatial bin at $X \approx 76''$ and $Y \approx -997''$, corresponding to an interplume area. Both the iron DEM (orange line) and the Fe \& Si DEM (blue line) are displayed. For this particular spatial bin, 13 Fe and 3 Si lines match the criteria described in Section~\ref{subsub:reliability_dem}, ensuring that the DEM is well constrained. The horizontal error bars represent the temperature resolution of the inversion and the vertical error bars are the uncertainties arising from the intensity measurements~\citep{hannah2012}. Each star represents the quantity $\xi_{\rm p}(T_{\rm t}) I_{\rm pred}/I_{\rm obs}$ computed for the Fe \& Si DEM (see Section~\ref{subsub:silicon_dem} for details). There is good agreement between the Fe and the Fe \& Si DEMs, indicating that both Fe and Si have a similar FIP bias. Also notable is the very good agreement between Fe and Si in the temperature range where \ion{S}{8} forms (green star). \label{fig:dem_ex}}
\end{figure*}

The $f_{\rm Fe}/f_{\rm Si}$ ratio is close to unity, which implies that Fe and Si have a similar FIP bias. This motivates our computing a Fe \& Si DEM using the Fe and Si lines for each spatial bin. We used Equation~(\ref{eq:int_fe_dem}) for the Fe lines. For the Si lines, we used a similar equation but with $A_{\rm{Fe}}^{\rm p}$ in place of $A_{\rm{Si}}^{\rm p}$. This allows more lines to be used in the inversion process, thereby providing a more robust DEM estimation. The $\xi_{\rm p}$ DEM, inferred using photospheric abundances, is now given by 
\begin{equation}
\label{eq:dem_fip}
\xi_{\rm p}(T_{\rm e}) = f_{(\rm Fe\ \&\ Si)} \it \xi_{\rm c}(T_{\rm e}),
\end{equation}
where $f_{(\rm Fe\ \&\ Si)} \equiv f_{\rm Fe} = f_{\rm Si}$. Figure~\ref{fig:dem_ex} displays a typical example of both Fe and Fe \& Si DEMs. The stars represent the quantity $\xi_{\rm p}(T_{\rm t}) I_{\rm pred}/I_{\rm obs}$ for the Fe, Si, and S lines, computed using the Fe \& Si DEM. $T_{\rm t}$ is the DEM weighted formation temperature, given by
 
\begin{equation}
\label{eq:temp_tt}
\log T_{\rm t} = \frac{\int G_{ji}(T_{\rm e})\xi_{\rm p}(T_{\rm e})\log T_{\rm e} dT_{\rm e}}{\int G_{ji}(T_{\rm e})\xi_{\rm p}(T_{\rm e})dT_{\rm e}}.
\end{equation}  
The quantity $\xi_{\rm p}(T_{\rm t}) I_{\rm pred}/I_{\rm obs}$ illustrates the match between the inferred DEM and the intensity data.  

The Fe and the Fe \& Si DEMs are very similar, though there are some differences. We attribute these, in part, to the different number of lines used and thus to the different temperature coverage. In general we find good agreement between the predicted and observed intensities, as illustrated in Figure~\ref{fig:dem_ex} by the closeness of most of the stars to the DEM curve.

\begin{figure*}
\begin{center}
\epsscale{0.9}
\includegraphics[angle=270, scale = 0.57]{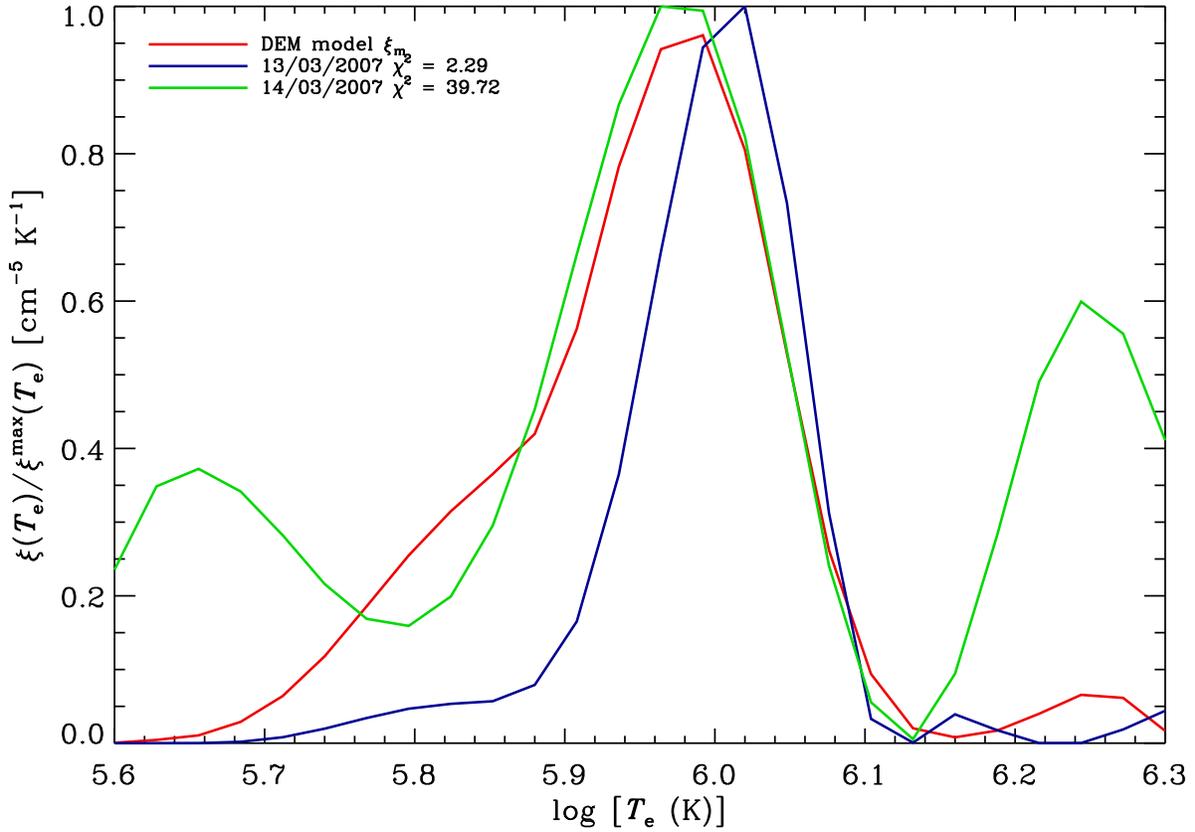}
\end{center}
\caption{The normalized Fe \& Si DEM model $\xi_{\rm m}$ (red line) computed for the first data set, where there is little contamination from high temperature emission. The blue line shows an example of an inferred, normalized DEM close to the model ($\chi^2 = 2.30$), located at $X\approx-36$ and $Y\approx-1032$. The green line shows an inferred normalized DEM with a significant contribution from quiet Sun plasma, leading to a $\chi^2 = 39.54$, for the spatial bin at $X\approx96$ and $Y\approx-1001$.  According to the $\chi^2$ test described in Section~\ref{subsub:silicon_dem}, we rejected this spatial bin from our analysis. \label{fig:dem_model}}
\end{figure*}

The main contribution to the DEM comes from plasma between $\log T_{\rm e} = 5.9$ and 6.1, which is the characteristic temperature range for plume and interplume plasma. There is also a small contribution at high temperature, between $\log T_{\rm e} = 6.2$ and 6.3, coming from quiet Sun material lying along the LOS. This signal complicates the interpretation of the results~\citep[][]{hahn2011} and can distort the abundance measurements as the quiet Sun is subject to a small FIP effect~\citep[][]{feldman2003, feldman2007}. To avoid this contamination, we developed an additional test, based on Pearson's $\chi^2$ test, that rejects spatial bins for which this quiet Sun contamination is excessive. First, we generated a Fe \& Si DEM model from the DEMs inferred from the first data set. For this particular observation, there is little or no contamination from the quiet Sun. A DEM model, $\xi_{\rm m}$, is computed by averaging the normalized inferred DEMs $\xi_{\rm p}/\xi_{\rm p}^{\rm max}$ over all gas at $Y=-1007''$. This average DEM is plotted in Figure~\ref{fig:dem_model}. Next, for each spatial bin of each data set we applied Pearson's $\chi^2$ test in order to evaluate how much the inferred Fe \& Si DEMs differ from the model. We then reject Fe \& Si DEMs with a significant high temperature contribution. For this, we computed Pearson's cumulative test statistic

\begin{equation}
\chi^2 = \sum_{j=0}^{N_{\rm bin}} \frac{[\xi_{\rm p}(T_{\rm e}^j)/\xi_{\rm p}^{\rm max} - \xi_{\rm m}(T_{\rm e}^j)]^2}{\xi_{\rm m}(T_{\rm e}^j)},
\end{equation}          
where $j$ indexes a point on the temperature grid and $N_{\rm bin} = 26$ is the number of points on this grid. If the inferred DEMs are only affected by normally distributed random errors, then Pearson's cumulative test statistic $\chi^2$ is equivalent to a statistical chi-squared with $N_{\rm bin}$ degrees of freedom. Spatial bins are rejected if $\chi^2$ is too large. We define the threshold for this to be $\chi^2 > 38.885$. The probability of have such a large $\chi^2$ for a statistical 26-degrees-of-freedom, chi-squared distribution is less than 5\%. That implies a 95\% chance that there is significant quiet Sun contamination in the spatial bin.      

\begin{figure*}
\begin{center}
\includegraphics[scale = 0.9, trim=450 150 700 90]{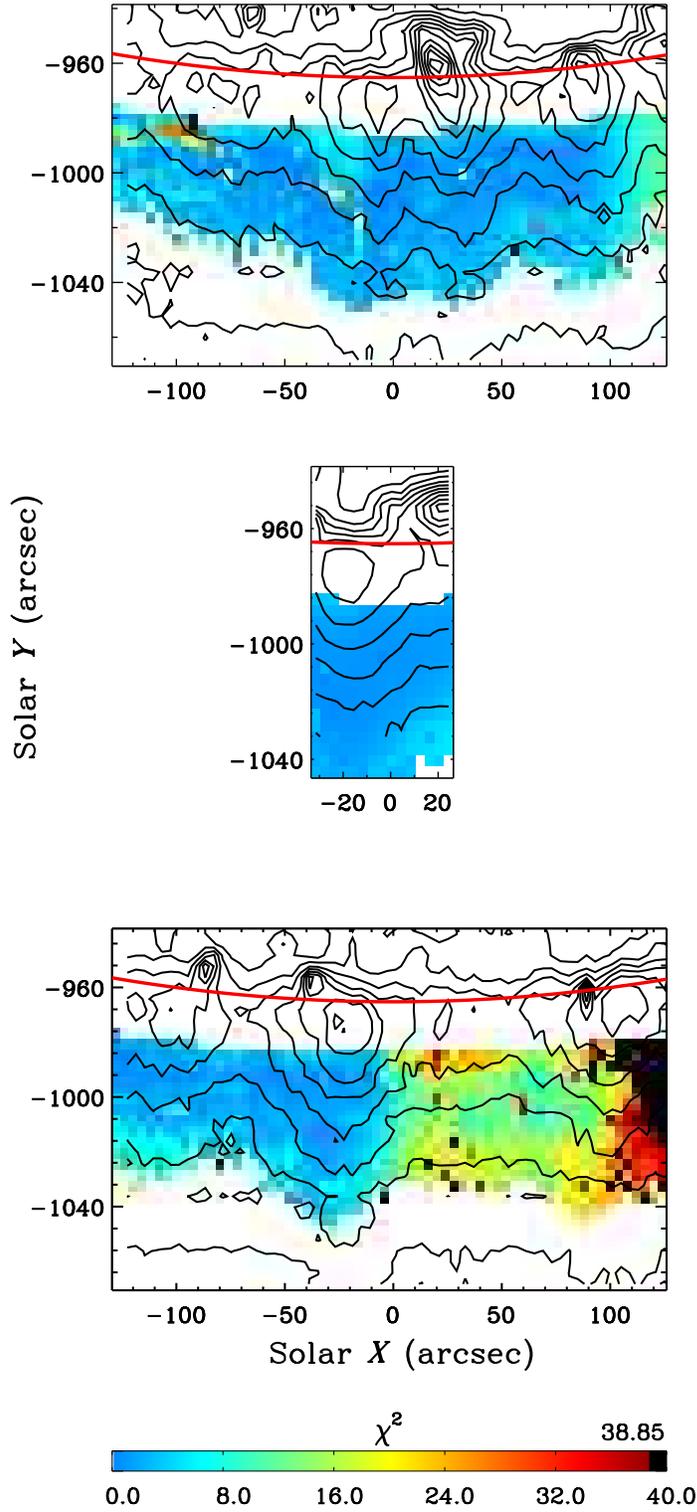}
\end{center}
\caption{Maps of Pearson's cumulative test statistic $\chi^2$ for each data set, in the same order as in Figure~\ref{fig:density}. The black regions correspond to spatial bins where $\chi^2 > 38.885$ and which have been rejected due to excessive quiet Sun material. The last observation is the most contaminated by high temperature emission, as can be seen on the right hand side of the map. The black contours are as in Figure~\ref{fig:density} and the red curve as in Figure~\ref{fig:si_maps}.\label{fig:chi2_maps}}
\end{figure*}

Figure~\ref{fig:chi2_maps} shows the $\chi^2$ maps for each of the three data sets. The first and second maps exhibit very homogeneous $\chi^2$ values, with most of the values falling between 0 and 5, indicating that the observations are of relatively pure coronal hole plasma. However, the last observation clearly shows that some of the inferred DEMs contains significant quiet sun contamination. These spatial bins have been rejected. 

\subsection{Computing the Fe \& Si to S FIP-bias ratio}
\label{subsub:fip_bias}
 
In order to derive the relative FIP bias, we compared the observed S~VIII intensities to the intensities predicted using the inferred Fe \& Si DEM $\xi_{\rm p}$. From this DEM, derived in Section~\ref{subsub:silicon_dem}, the predicted intensities are given by
\begin{equation}
\label{eq:int_s_pred}
I^{\rm S}_{\rm pred} = \it f_{(\rm Fe\ \&\ Si)} \frac{A_{\rm{S}}^{\rm p}}{\rm 4\pi} \it \int_{0}^{+\infty} G_{ji}(T_{\rm e}, n_{\rm e})  \xi_{\rm p}(T_{\rm e}) d\log T_{\rm e}. 
\end{equation}
The observed S VIII 198.55 \AA\ intensities are given by
\begin{equation}
\label{eq:int_s_obs}
I^{\rm S}_{\rm obs} = f_{\rm S}\it \frac{A_{\rm{S}}^{\rm p} }{ \rm 4\pi} \it \int_{0}^{+\infty} G_{ji}(T_{\rm e}, n_{\rm e}) \xi_{\rm p}(T_{\rm e}) d\log T_{\rm e}, 
\end{equation}
where $f_{\rm S}$ is the sulfur FIP bias. The ratio of the predicted-to-observed S~VIII~198.55~\AA\ intensities is simply
\begin{equation}
\label{eq:ratio_S}
\frac{I^{\rm S}_{\rm pred}}{I^{\rm S}_{\rm obs}} = \frac{f_{(\rm Fe\ \&\ Si)}}{f_{\rm S}}.
\end{equation}
Using this for each spatial bin of our three observations, we have generated maps of $f_{(\rm Fe\ \&\ Si)}/f_{\rm S}$ which we discuss in Section~\ref{sec:fip_maps}.
  
\section{Results}
\label{sec:fip_maps}

\begin{figure*}
\begin{center}
\includegraphics[scale = 0.9, trim=450 150 700 90]{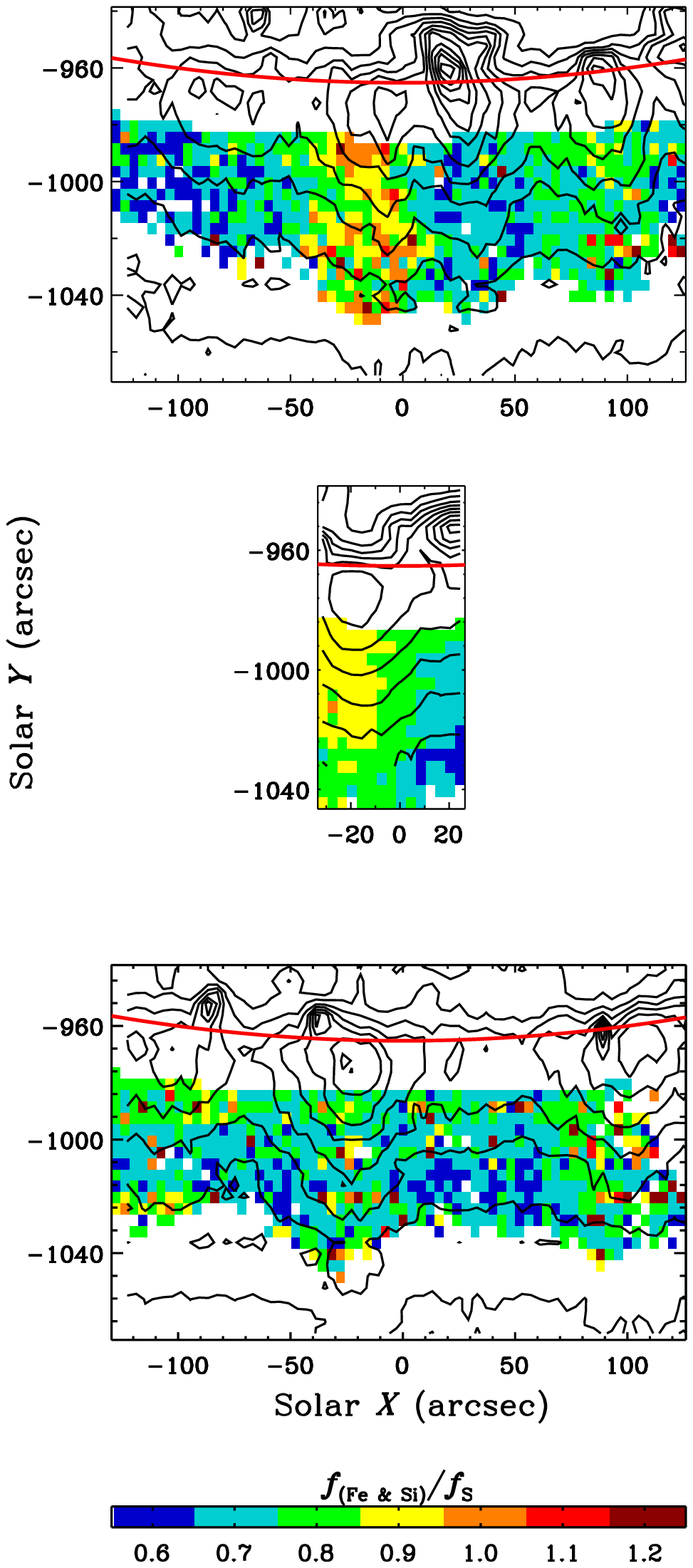}
\end{center}
\caption{Same as Figure~\ref{fig:si_maps}, but for the FIP-bias ratio $f_{(\rm Fe\ \&\ Si)}/f_{\rm S}$.  \label{fig:s_maps}}
\end{figure*}

\begin{figure*}
\begin{center}
\includegraphics[scale = 0.9, trim=450 200 700 90]{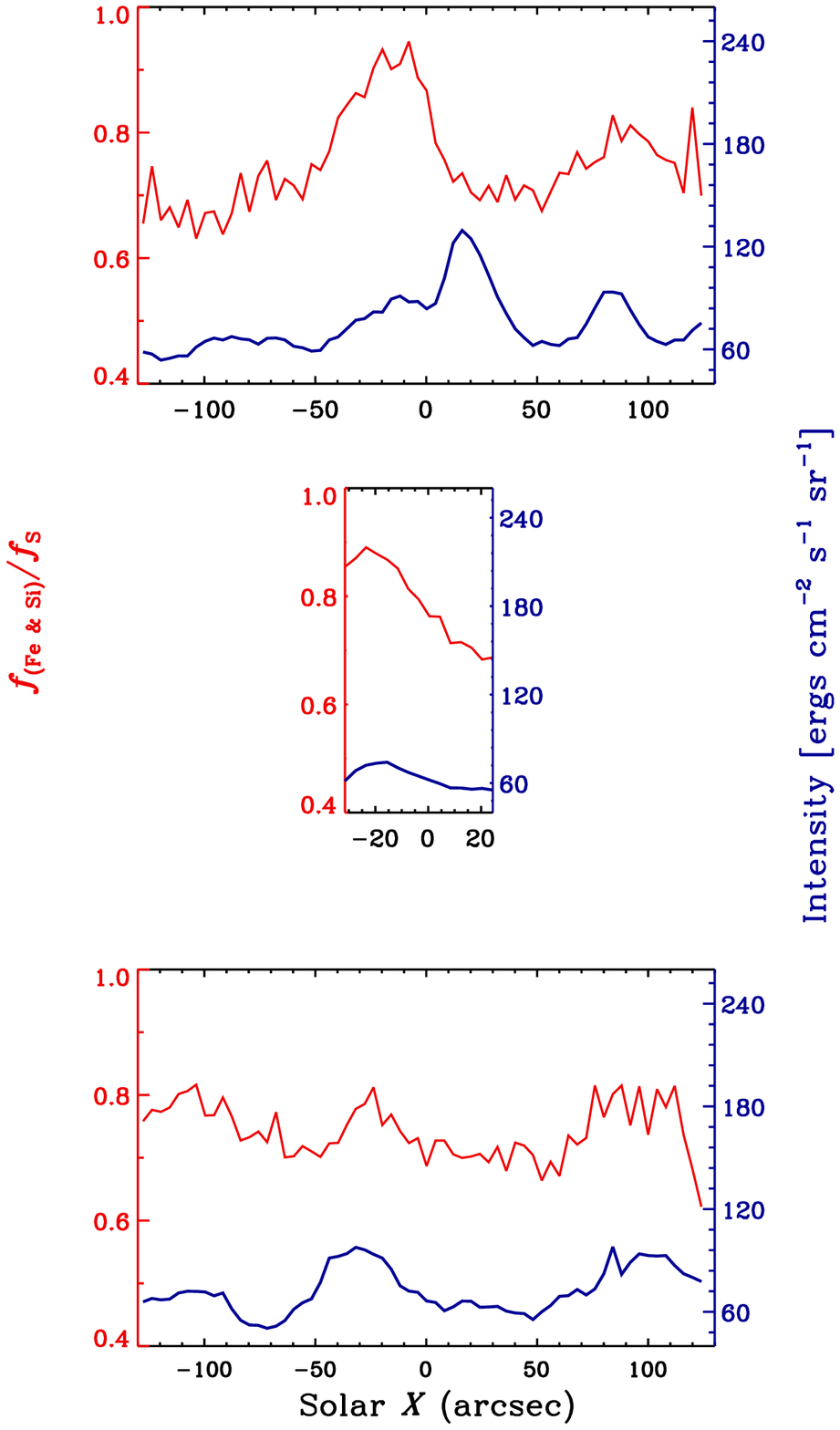}
\end{center}
\caption{Same as Figure~\ref{fig:profiles_fe}, but for the FIP-bias ratio $f_{(\rm Fe\ \&\ Si)}/f_{\rm S}$. \label{fig:profiles_s}}
\end{figure*}

Figure~\ref{fig:s_maps} shows the $f_{(\rm Fe\ \&\ Si)}/f_{\rm S}$ FIP-bias-ratio maps for the three data sets. The corresponding profiles, computed for the maps given in Figure~\ref{fig:s_maps} by averaging the spatial bins over the Y
direction, are displayed in the Figure~\ref{fig:profiles_s}. Also shown in Figure~\ref{fig:profiles_s} are the intensity profiles observed using the~\ion{Fe}{10}~184.54~\AA\ line. Table~\ref{Tab:Results} lists the identified plume and interplume regions in each observation; the approximate $X$ location of their intensity maximum or minimum, repsectively; and their approximate FIP-bias ratios. 

The top panel of Figure~\ref{fig:s_maps} shows $f_{(\rm Fe\ \&\ Si)}/f_{\rm S}$ for the first data set, while the corresponding profiles are shown in the top panel of Figure~\ref{fig:profiles_s}. Two interplume regions are visible at $X \approx -50''$ and $60''$. Three plumes are easily identified in this particular observation, located respectively at $X \approx -15''$, $15''$ and $85''$. These coincide with the intensity enhancement seen in the contour lines of Figure~\ref{fig:s_maps} and the intensity profiles of Figure~\ref{fig:profiles_s}. There is a suggestion of a fourth plume, located at $X \approx -85''$ and visible on the EIT field of view of Figure~\ref{fig:density}. 

The two interplume regions present a FIP-bias ratio of $\approx 0.7$. Conversely, in the plume regions the FIP-bias ratio varies from plume to plume. The plume located at $\approx -15''$ exhibits the largest relative FIP-bias ratio, with values close to 0.95. The plume visible at $\approx 85''$, shows an enhancement of about 0.8. On the other hand, the brightest plume, detected at $\approx 15''$, has a FIP-bias ratio of about $0.7$, similar to what we measure in the interplume regions. We save discussion of the plume at $\approx -85''$ until later.  

The middle panels of Figure~\ref{fig:s_maps} and~\ref{fig:profiles_s} display the $f_{(\rm Fe\ \&\ Si)}/f_{\rm S}$ measured for the second data set. Even though the field of view is smaller, we still clearly observe the plume located around $X \approx-20''$. The brightest plume, detected at around $X\approx 15''$ in the top panel, has faded away, and an interplume region is now visible at $X\approx 10''$. The plume still exhibits a FIP-bias ratio of about $0.9$. This is slightly smaller than in the first data set, but still larger than the FIP-bias ratio of the interplume region, which is about 0.7, similar to the results from the first data set. 

The bottom panels of Figure~\ref{fig:s_maps} and~\ref{fig:profiles_s} present the last data set, collected $\approx$ 24 hours after the first. Two plumes are still clearly visible, now located at $X \approx -30''$ and $100''$. We also find a suggestion of a third plume at $X \approx -100''$, which is rather fainter than the other two. This may be the plume from the first data set at $X \approx -85^{\prime\prime}$. We attribute the shifts in the positions of all the observed structures to a combination of the solar rotation and movement of the field lines caused by other possible processes. Interplume regions are detected around the $X \approx -70''$ and $50''$, and continue to exhibit FIP-bias ratios of about 0.7. The plume around $X \approx -30''$, which displayed the largest enhancement in the first two data sets, now exhibits a reduced FIP-bias ratio of about 0.8. This is particularly clear when looking at the profiles in Figure~\ref{fig:profiles_s}. Also notable is that all three of the remaining plumes have a similar FIP-bias ratio of about $0.8$.  

To summarize this section, we find that interplume gas exhibits a constant FIP-bias ratio over the duration of the observations. Conversely, over the $\approx$ 24 hour observation period the time evolution of the FIP-bias ratio differs dramatically from plume to plume. The initially brightest plume at $X \approx 15''$, which appears to originate from a bright point, rapidly fades away, becoming an interplume region, and does not show any evolution in its FIP-bias ratio, with an initial FIP-bias ratio similar to that of the interplume regions. This is to be contrasted with the behavior of the other three plumes. The plume initially at $X \approx -15''$ shows the largest initial FIP-bias ratio, but the value decreases over time, approaching, though not reaching, that of the interplume gas. These changes occur despite the apparently nearly constant brightness and density of this structure. The plume initially at $X \approx -80''$ shows a small increase in its FIP-bias ratio over time, increasing from a value initially close to that of interplume gas. Lastly, the plume initially at $X\approx 85''$ exhibits a nearly constant FIP-bias ratio, which is slightly higher than that of the interplume gas. 

\section{Systematic Uncertainties}
\label{res_accuracy}

There are several source of uncertainties involved in our analysis.  In this section we attempt to identify these sources and discuss their possible effects on our results. 

\subsection{Instrumental}
\label{sub:instrumental}
Instrumental factors, such as the calibration of the EIS instrument could produce a systematic bias in the results. \citet{culhane2007} estimated the absolute calibration of the EIS spectrometer to be around 25\% for each of the CCD cameras on the instrument. However the relative calibration error within a given camera or between the two cameras is expected to be much smaller. Thus we expect this error essentially cancels out in our analysis. Additional instrumental issues could arise due to degradation effects. However, we used data from 2007, at the beginning of the EIS observations period, when there had not yet been any significant degradation~\citep{delzanna2013}. Thus no significant instrumental uncertainties are expected in our results. 

\subsection{Atomic data}
\label{sub:atomic}

Errors associated with the atomic data required to interpret the observed intensities are another potential source of systematic uncertainty, but such errors are difficult to estimate~\citep{feldman2009}. One source of error could be in the line emissivities due to uncertainties in the excitation rate coefficients, transition rates, and incompleteness of the cascade contributions.~\citet{delzanna2013} has carried out a detailed study of solar observations compared to model predictions and identified a number of lines where the model line emissivities are likely to be in error.  We have used their results as a guide and avoided all those lines. Hence, we expect the line emissivity uncertainties to be minimal for the lines used in the present study.

Another source of atomic physics uncertainty lies with the ionization balance calculations. There has been extensive experimental and theoretical work to improve these data as summarized in~\citet{bryans2009}. Our results here used the ionization balance calculations of CHIANTI version 7.1 which has updated the data of~\citet{bryans2009} by including state-of-the-art recombination rate coefficients for isoelectronic sequences published after their work.  Still, in order to estimate the effects of the new ionization balance calculations on our measurements, we repeated the entire analysis using the data from~\citet{bryans2009}. We did not find any substantial change from the results presented here. The difference between the two FIP-bias-ratio maps never exceeds 5\%, and the relative variations of the FIP-bias ratio between plumes and interplumes is unchanged.

\subsection{Intensity ratios: the isothermal approximation}
\label{sub:int_rat}
 
\begin{figure*}
\begin{center}
\epsscale{0.8}
\includegraphics[angle=270, scale = 0.7]{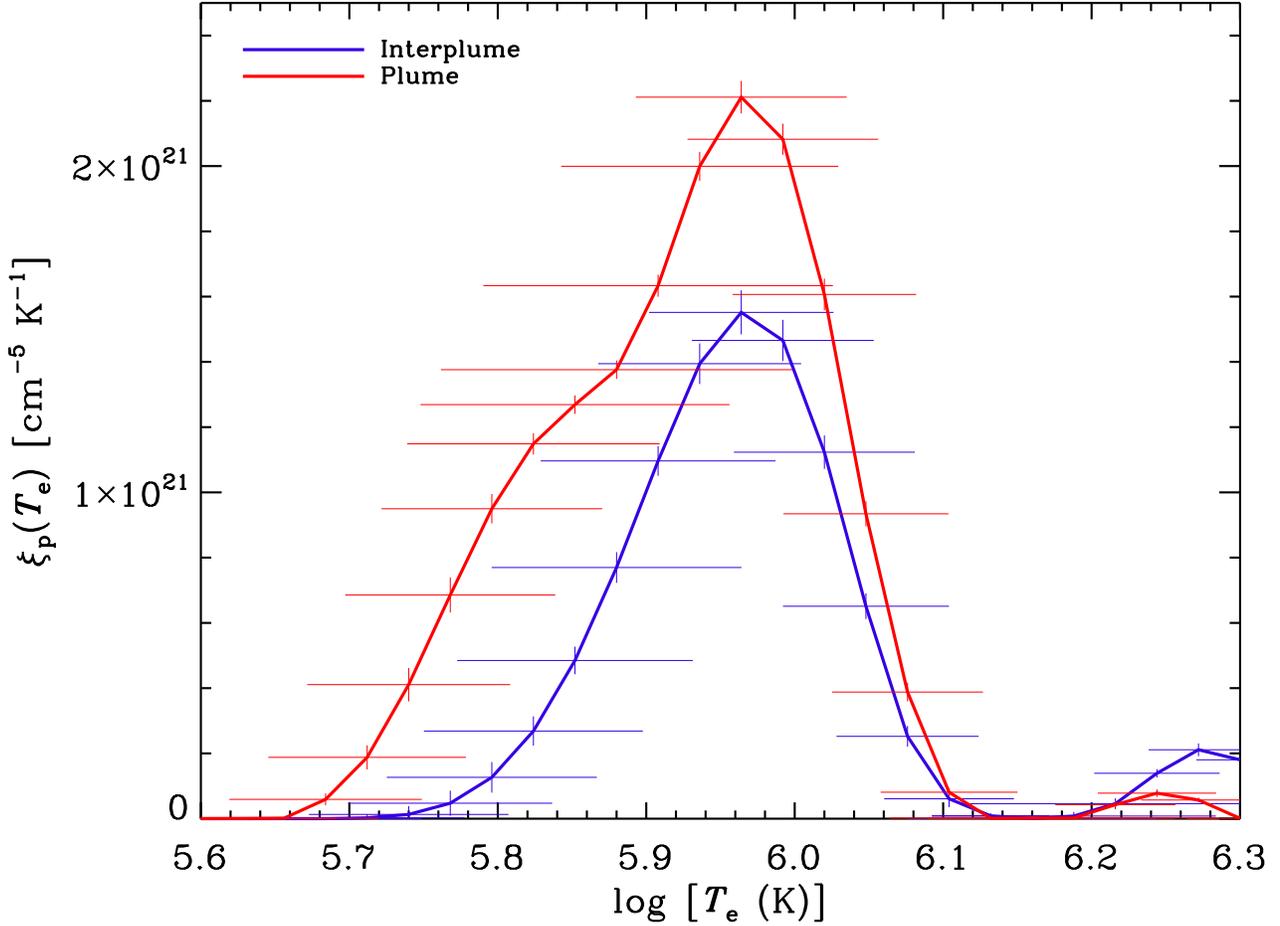}
\end{center}
\caption{Typical plume (red line) and interplume (blue line) Fe \& Si $\xi_{\rm p}$ DEMs. The plume DEM is computed for the spatial bin located at $X = -8''$, while the interplume is at $X = 57''$. The solar-$Y$ coordinate in the both cases is $Y=-997''$. The horizontal error bars represent the temperature resolution of the inversion and the vertical error bars are the uncertainties arising from the intensity measurements. \label{fig:dem_plume_ipp}}
\end{figure*}

The intrinsic differences between the plume and interplume DEMs might introduce a systematic bias in the inferred FIP-bias ratio. Figure~\ref{fig:dem_plume_ipp} displays typical inferred Fe \& Si $\xi_{\rm p}$ DEMs for plume and interplume gas. Both $\xi_{\rm p}$ functions peak around $\log T_{\rm e} = 5.95$, but the interplume peak is weaker than the plume, due to its lower density. The plume DEM also exhibits a higher contribution from plasma between $\log T_{\rm e} = 5.8-5.9$, coming from the cooler part of the plume plasma. 

We find that the DEMs measured in both plumes and interplumes are narrow in temperature. Within the  Gaussian approximation, the full width at half maximum (FWHM) is $\Delta \log T_{\rm e} \approx 0.15$ for the interplume regions, and $\Delta \log T_{\rm e} \approx 0.2$ in plume regions. These are small values, considering the temperature resolution of the DEM inversion. Indeed,~\citet{landi2010} and~\citet{guennou2012a} have demonstrated that plasmas with a Gaussian-DEM FWHM of $\Delta \log T_{\rm e} \lesssim 0.10$ are consistent with the gas being isothermal. This is due to the presence of systematic and random errors, which limit the intrinsic temperature resolution of the DEM inversion.

The narrow widths of the DEMs for both the plumes and interplumes enables us to use the isothermal approximation in order to estimate the error on the absolute FIP-bias-ratio results. This also allows us to test the robustness of the relative variation in the inferred FIP-bias ratios. For this we use the \ion{Si}{7}~275.39~\AA\ and \ion{S}{8}~198.55~\AA\ lines, which are particularly well suited for FIP-bias-ratio measurements of coronal hole plasma, since most of their line emission is from the temperature ranges typical of coronal holes \citep{feldman2009}.  

In the isothermal approximation, the observed intensity of the \ion{Si}{7}~275.39~\AA\ line for a plasma at a given temperature $T_{\rm e}$ and density $n_{\rm e}$ is 
\begin{equation}
\label{eq:int_si_iso}
I^{\rm Si}_{\rm obs} =\it f_{\rm Si} \frac{A_{\rm{Si}}^{\rm p} }{ \rm 4\pi} G_{ji}^{\rm Si}(T_{\rm e}, n_{\rm e}) \times \rm EM,
\end{equation}
where $G_{ji}^{\rm Si}$ is the Si contribution function and EM is the emission measure, which is defined using Equation~(\ref{eq:int_ne}) as
\begin{equation}
\label{eq:em}
{\rm EM} = \int n_{e}^2 dl.
\end{equation}
In the same way, the \ion{S}{8}~198.55~\AA\ intensity can be written
\begin{equation}
\label{eq:int_s_iso}
I^{\rm S}_{\rm obs} =\it f_{\rm S} \frac{A_{\rm{S}}^{\rm p} }{\rm 4\pi} G_{ji}^{\rm S}(T_{\rm e}, n_{\rm e}) \times \rm EM,
\end{equation}
where $G_{ji}^{\rm S}$ is the S contribution function.

The intensity ratio of the \ion{Si}{7}~275.39~\AA\ to the \ion{S}{8}~198.55~\AA\ lines gives
\begin{equation}
\label{eq:ratio_int}
\frac{I^{\rm S}_{\rm obs}}{I^{\rm Si}_{\rm obs}} = \frac{f_{\rm S}}{f_{\rm Si}} \frac{A_{\rm{S}}^{\rm p} }{A_{\rm{Si}}^{\rm p}} \frac{G_{ji}^{\rm Si}(T_{\rm e}, n_{\rm e})}{G_{ji}^{\rm S}(T_{\rm e}, n_{\rm e})},
\end{equation}
and thus the FIP-bias ratio $f_{\rm Si}/f_{\rm S}$ can be calculated using
\begin{equation}
\label{eq:fip_iso}
\frac{f_{\rm Si}}{f_{\rm S}} = \frac{A_{\rm{S}}^{\rm p} }{A_{\rm{Si}}^{\rm p}} \frac{G_{ji}^{\rm Si}(T_{\rm e}, n_{\rm e})}{G_{ji}^{\rm S}(T_{\rm e}, n_{\rm e})} \frac{I^{\rm Si}_{\rm obs}}{I^{\rm S}_{\rm obs}}.
\end{equation}
In order to estimate the FIP-bias ratio using Equation~(\ref{eq:fip_iso}), we need to calculate the contribution function for the \ion{Si}{7}~275.39~\AA\ and \ion{S}{8}~198.55~\AA\ lines. For this, the density has been estimated using the \ion{Fe}{9} line intensity ratio as described in Section~\ref{subsec:density}. Additionally, for each spatial bin, we estimated $T_{\rm e}$ as the temperature corresponding to the peak of the DEM.

\begin{figure*}
\begin{center}
\includegraphics[scale = 0.9, trim=450 150 700 90]{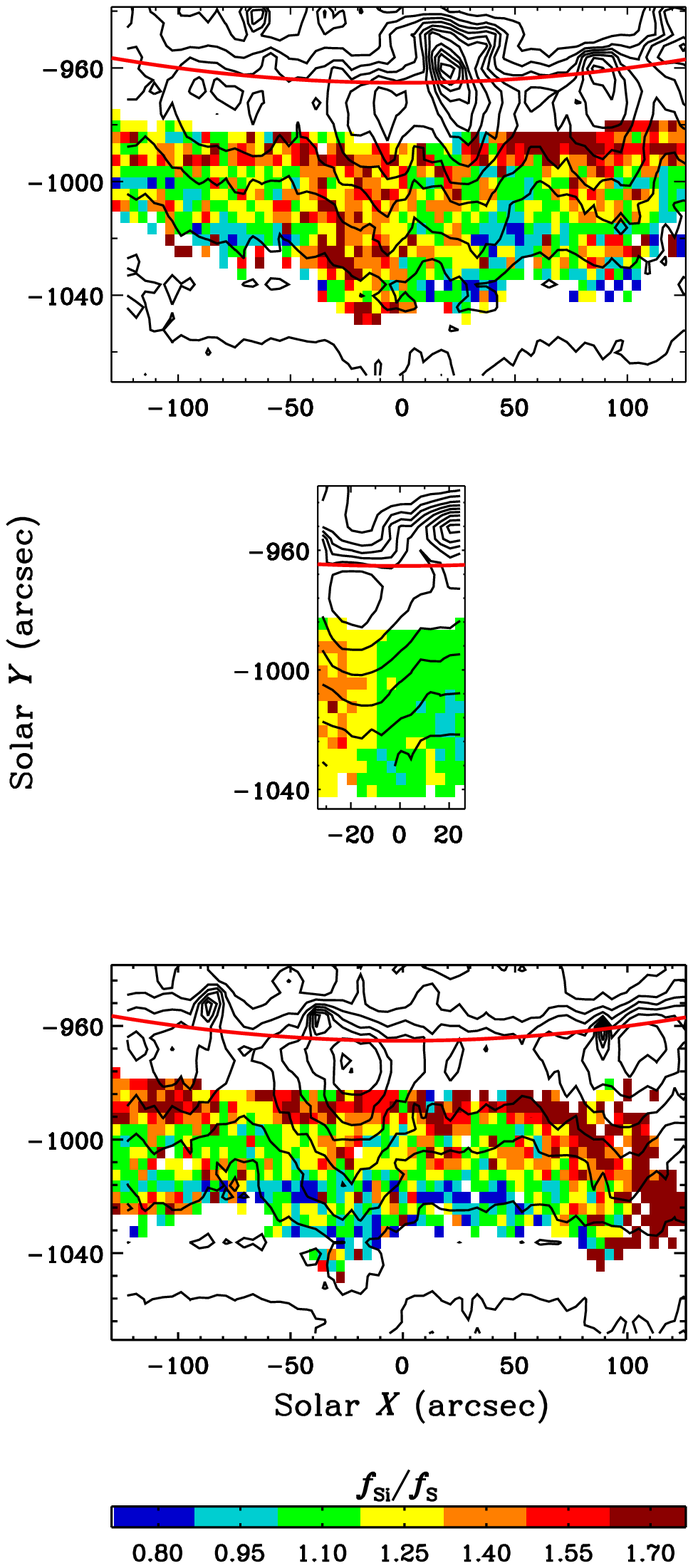}
\end{center}
\caption{Same as Figures~\ref{fig:si_maps} and~\ref{fig:s_maps}, but for the FIP-bias ratio $f_{\rm Si}/f_{\rm S}$ using an isothermal analysis. These maps have been obtained using the line pairs diagnostic \ion{Si}{7}~275.39~\AA\ and \ion{S}{8}~198.55~\AA\ \label{fig:int_ratio_maps}.}
\end{figure*}

\begin{figure*}
\begin{center}
\includegraphics[scale = 0.9, trim=450 200 700 90]{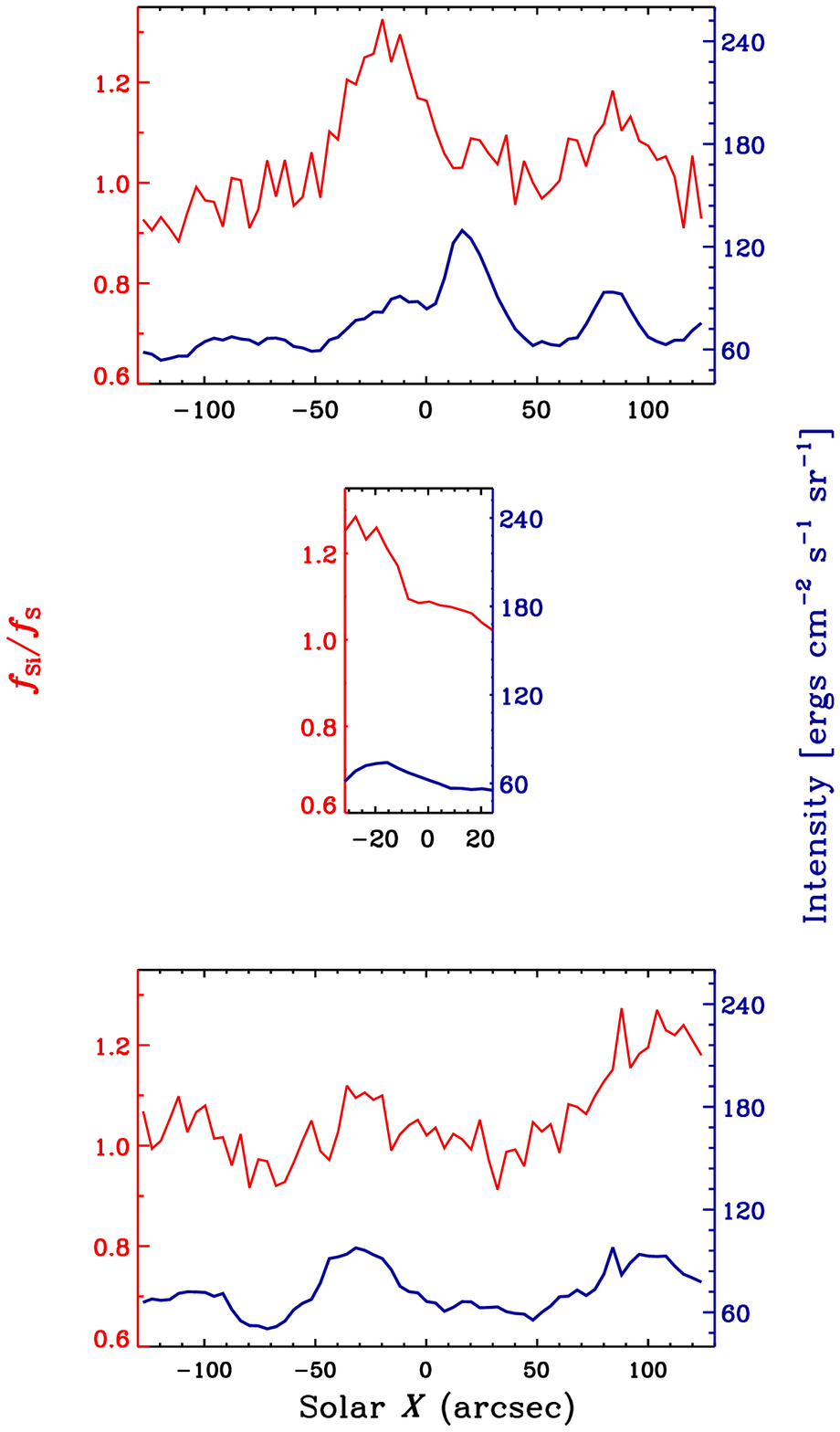}
\end{center}
\caption{Same as Figures~\ref{fig:profiles_fe} and~\ref{fig:profiles_s}, but for the FIP-bias ratio $f_{\rm Si}/f_{\rm S}$ using an isothermal analysis. These curves have been obtained using the line pairs diagnostic \ion{Si}{7}~275.39~\AA\ and \ion{S}{8}~198.55~\AA\ . \label{fig:profiles_ratio}}
\end{figure*}

Figures~\ref{fig:int_ratio_maps} and~\ref{fig:profiles_ratio} respectively show the resulting FIP-bias-ratio maps and the profiles of the FIP-bias ratio $f_{\rm Si}/f_{\rm S}$ for our three data sets. Table~\ref{Tab:Results_iso} lists the identified plume and interplume regions in each observation; the approximate $X$ location of their intensity maximum or minimum, respectively; and their approximate FIP-bias ratios. 

The top panels of Figures~\ref{fig:int_ratio_maps} and~\ref{fig:profiles_ratio} show that the two interplumes regions, located at $X \approx -50''$ and $60''$, have FIP-bias ratios of $\approx 1.05$ and $0.97$, respectively. The two plumes located at $X \approx -15''$ and $85''$ exhibit higher FIP-bias ratios of $\approx 1.32$ and 1.12, respectively. Conversely, the plumes at $X \approx -80''$ and $15''$ have FIP-bias ratios slightly smaller than those of the interplumes regions. In general, these relative results are similar to those of the DEM analysis, though the absolute values of the FIP-bias ratios are different. 

In the middle panels, corresponding to the second data set, the plume is now located at $X \approx -20''$ and shows a FIP-bias ratio of $\approx 1.23$. The newly formed interplume region at $X \approx 10''$ exhibits a FIP-bias ratio $\approx 1.12$, which is actually higher than the now-faded-plume results of the first data set. Thought the differences between the two regions are less marked than in the DEM analysis, the general trend is similar.

The bottom panels of Figures~\ref{fig:int_ratio_maps} and~\ref{fig:profiles_ratio} correspond to the third data set. The FIP-bias ratio of the plume located at $X \approx -100''$ has increased slightly, similar to what was seen in the DEM analysis; however, the ratio is not significantly different from that of the interplume results. For the plume at $X \approx -30''$, we find a decreasing trend in the FIP-bias ratio over the course of the three observations, similar to what was seen in the DEM analysis. Lastly, the FIP-bias ratios of the plume located at $X \approx 100''$ is about 1.25, higher than that of the interplume gas in the first observation. However, this plume exhibits a different behavior than was inferred from using the DEM analysis. In the isothermal case the FIP-bias ratio increases, whereas in the DEM analysis it remained nearly constant.

The isothermal analysis, similar to that of the DEM, finds that interplume gas exhibits a roughly constant FIP-bias ratio over the duration of the observations. Additionally, over the $\approx 24$ hour observation period the time evolution of the FIP-bias ratio varies from plume to plume. Despite the minor differences between the isothermal and DEM analyses of the plumes, this strengthens the conclusions drawn from the DEM results. However, we still need to address the uncertainties in the inferred FIP-bias ratios.  

We used a Monte-Carlo procedure to estimate the uncertainty associated with the isothermal measurements of the FIP-bias ratio $f_{\rm Si}/f_{\rm S}$. For this we took into account the errors on the intensity, density, and temperature measurements. In order to estimate the effect from each of these uncertainties, we randomly varied one of the parameters, using a normal distribution and the associated 1$\sigma$ error, while keeping the other two parameters fixed. The modified parameter is used to estimate the FIP-bias ratio, and the operation is repeated 500 times. The impact of the uncertainty in each parameter on the FIP-bias ratio is deduced by computing the standard deviation of these 500 different results. The individual uncertainties associated with the intensity, density, and temperature measurement errors are reported in Table~\ref{Tab:Results_iso}.

\subsubsection{Uncertainty due to the Intensity and Density Measurements}
\label{subsub:unc_int}

The statistical uncertainties in the intensities have been computed as described in Section~\ref{subsec:lines}. There may also be some systematic uncertainties related to the absolute calibration of the EIS spectrometer (see Section~\ref{sub:instrumental}). Since the FIP-bias-ratio calculation takes the ratio of the \ion{Si}{7}~275.39~\AA\ and \ion{S}{8}~198.55~\AA\ line intensities, this systematic error essentially cancels out and only the small relative calibration error remains.  The impact of this relative error on the FIP-bias-ratio estimation is expected to be insignificant.

We find that the statistical errors involved in the intensity measurements give rise to a $1\sigma$ uncertainty in the FIP-bias ratio of between 0.08 and 0.12, varying with the location in the observations. In general, this is smaller than the observed FIP-bias-ratio differences. Hence we conclude that uncertainties in the intensity measurements cannot explain the observed structures.

The uncertainty in the density has been estimated using the computed $1\sigma$ errors on the \ion{Fe}{9}~188.50~\AA\ and 189.94~\AA\ lines intensities. As described in Section~\ref{subsec:density}, the density diagnostic assumes that the plasma is isothermal at the temperature corresponding to the maximum of the \ion{Fe}{9}~188.50~\AA\ and 189.94~\AA\ contribution functions, which peak at about $\log T_{\rm e} \approx 5.9$. This temperature is essentially indistinguishable from the peak temperature of the DEMs for both plumes and interplumes. Hence, we expect any small systematic errors due to the temperature used to be insignificant. Additional systematic uncertainties could arise due to errors with the excitation rate coefficients, transition rates, and incompleteness of the modeled cascade contributions to the \ion{Fe}{9}~188.50~\AA\ and 189.94~\AA\ line intensities. Fortunately, the contribution functions for the \ion{Si}{7}~275.39~\AA\ and \ion{S}{8}~198.55~\AA\ are only slightly dependent on the density. As a result, systematic uncertainties related to the density measurements are expected to only have a small impact on the FIP-bias ratio calculations. We therefore only consider the statistical error related to the intensity measurements. 

We find that errors in the density measurements give rise to a $1\sigma$ uncertainty in the FIP-bias ratio of less than 0.02 for both plume and interplume gas. This is smaller than most of the observed FIP-bias-ratio differences. Hence we conclude that uncertainties in the density measurements cannot explain the observed structures.

The quadrature sum of the uncertainties due to the intensity and density measurements gives a $1\sigma$ value of less than 0.12. Again, this is still smaller than most of the observed FIP-bias-ratio differences and strengthens our conclusion that the observed structure cannot be attributed to errors in the intensity or density measurements.

\subsubsection{Uncertainties due to the Temperature Measurements}
\label{subsub:unc_both}

The temperature uncertainty is taken from the work of~\citet{landi2010} and \citet{guennou2012a}. They showed that, due to DEM-inversion complications, the uncertainty of the inferred DEM-peak temperature is estimated to have a FWHM of $\Delta \log T_{\rm e} = 0.1$. This takes into account both the random and systematic uncertainties due to unknown atomic physics and instrumental calibration. However, as described in~\citet{landi2010} and~\citet{guennou2012a}, the contribution of the statistical uncertainties to this estimated temperature error is much smaller than that due to systematic causes. 

The FIP-bias uncertainty related to the temperature error is higher than that due to the intensity or density. The $1\sigma$ value ranges between 0.46 and 0.57. This is as expected, since the contribution functions are strongly dependent on the temperature.

At first glance, it appears that this uncertainty is large enough to wash out the structure seen in the three data sets. However, we find it unlikely that the temperature in plumes and interplumes would be mis-estimated in such a way as to artificially generate the observed structure. If such a systematic effect were at work, then all of the plumes would be likely to show a FIP-bias ratio different from that of the interplumes. What we see instead is that only some of the plumes differ from the interplumes in their FIP-bias ratio. Additionally, the DEM analysis effectively takes into account the temperature uncertainty in the isothermal analysis; but it shows a structure similar to the isothermal results. Hence we interpret the temperature uncertainty in the isothermal results to represent a systematic uncertainty in the absolute FIP-bias results. We also take it as an estimate for the systematic uncertainty in the DEM FIP-bias results. 

\subsubsection{Robustness of the Results}
\label{subsub:robustness}

Based on the above error analysis, we have a high degree of confidence in the robustness of our relative FIP-bias-ratio results.  The uncertainties due to the intensity and density measurements are too small to explain the observed relative differences; and the uncertainty due to the temperature is most likely to systematically affect the results. Taking this last uncertainty as an estimate for the systematic reliability of the absolute scale, we find that this $\approx \pm 0.5$ accuracy of the absolute scale can readily explain the difference seen between the DEM FIP-bias ratio $f_{(\rm Fe\ \&\ Si)}/f_{\rm S}$, which ranges between $\approx 0.7$ and 0.9, and the isothermal FIP-bias $f_{\rm Si}/f_{\rm S}$, which ranges between $\approx 1$ and 1.3.
  
\section{Discussion}
\label{sec:conclusions}

Over the 24 hour period studied, in the interplume regions the DEM-inferred FIP-bias ratio $f_{(\rm Fe\ \&\ Si)}/f_{\rm S}$ exhibit a nearly constant behavior. Conversely, the plume regions display a range of behaviors. One starts with a nearly interplume FIP-bias ratio, which then increases slightly with time. The plume with the highest initial FIP-bias ratio shows a decreasing ratio with time, though it never reaches interplume values. A third plume begins with an interplume FIP-bias ratio and then fades, becoming an interplume region with nearly the same initial FIP-bias ratio. The fourth plume shows a nearly constant FIP-bias ratio, but one which is great than the interplume value. We have a high degree of confidence in the robustness of these relative results. Our DEM findings are largely supported by our corresponding isothermal analysis. Thus, we conclude that the observed variations are most likely due to real compositional differences.

Building on our findings, the evolution of plumes with time could explain the discrepancies seen among previous studies~\citep{wilhelm1998, young1999, delzanna2003}. These past measurements may have observed plumes at different stages in their lifetime. Although only one plume in our study showed a clear elemental abundance variation over the $\approx 24$~hr period, our findings demonstrate that abundance evolution can occur in plumes. Clearly, though, the others remained essentially stable in time. More systematic studies are needed in order to determine what are the common compositional features of plumes.     

One possible explanation for the observed time evolution of the FIP-bias ratio is that it is related to the time at which the plume formed over the underlying bright point. Beam plumes are believed to originate from magnetic reconnection of emerging magnetic flux with the ambient unipolar magnetic field, thereby generating a bright point~\citep{wang1994, grappin2011}. Recently, \citet{raouafi2008} showed that jets that arise from magnetic reconnection are precursors of plumes. Previous studies have shown that the FIP bias is related to the confinement time of the plasma. Thus, the initial abundance enhancement in the plume may represent the FIP bias in the bright point at the time the jet occurred. Once the material confined in the bright point is ejected, photospheric material is then directly ejected through the open magnetic field lines, resulting in a subsequent change of the elemental abundances towards photospheric, as observed in our results. If this hypothesis is correct, FIP-bias ratio measurements, such as those presented here, can provide constraints on the confinement time of the plasma that is released when the plumes are formed. Such data may also provide important boundary conditions for models of the plumes formation process and the heating mechanism occurring at their base.

A possible future application of the approach laid out in this work is to be able to identify plume and interplume material in the solar wind. Comparing the present results to the available solar wind in situ measurements is complicated, because our FIP-bias-ratio measurements have significant systematic uncertainties. However, archival spectral observations from \textit{SOHO}/CDS are available. These data include emission lines from many more elements and ionization stages than do the EIS data. The CDS data would thereby enable the construction of better constrained DEMs, resulting in reduced uncertainties in the inferred FIP-bias-ratio measurements. We hope in a future work to use CDS data for such studies, since more plume studies are needed in order to better characterize any time evolution of elemental abundances and develop diagnostics so as to be able to identify the signature of plumes within the fast solar wind. 

At the moment, the Sun is in its active period, with minimally developed coronal holes and excessive quiet Sun contamination, thereby making present-day plume and interplume observations more difficult. However, future observations from spectrometers and in situ instruments on board \textit{Solar Orbiter} offer great opportunities for connecting polar plume and interplume observations to in situ solar wind measurements. Such observations are likely to help to answer the question of whether or not plumes contribute to the fast solar wind.

Lastly, our DEM analysis of the plume and interplume regions reveal a FIP-bias ratio $f_{(\rm Fe\ \&\ Si)}/f_{\rm S}$ that is systematically less than 1.0. This would imply either a depletion of the low-FIP elements or an enhancement of the sulfur. A depletion would suggest an inverse FIP effect, which to date has only been observed in the most active stars \citep[][]{feldman2001, brinkman2001}. Previous FIP bias measurements of coronal hole plasma have consistently found a low-FIP element enhancement~\citep[][]{feldman2005, feldman2007, grevesse2007}. As discussed in Section~\ref{res_accuracy}, we attribute the ratio here of less than 1 to a systematic error due to the uncertainty in the DEM shape. Future measurements could possibly reduce this error by using more lines covering a wider temperature range, thereby enabling a more accurate determination of the DEM.

\section{Summary} 
\label{sec:summary}  

In this work, we have described measurements of the FIP-bias ratio $f_{(\rm Fe\ \&\ Si)}/f_{\rm S}$ within plume and interplume plasma. We used spectroscopic observations over an $\approx$ 24 hour period in order to characterize the evolution of these structures with time. We have found that the FIP-bias ratio is essentially constant with time in interplumes but can vary with time in plumes. Future studies combined with corresponding in situ measurements of the solar wind may help to answer the question of whether or not plumes or interplumes contribute to the fast solar wind and may also provide constraints on possible formation and heating mechanisms of plumes.

\acknowledgments
 
The authors would like to thank the technical staff at the Computer Science Department of the Institut d'Astrophysique Spatiale, Orsay, France, for allowing us to use their computational facilities. This work was supported in part by the NSF Division of Atmospheric and Geospace Sciences SHINE program grant AGS-1060194 and the NASA Solar Heliospheric Division Living with a Star Science Program grant NNX15AB71G. Hinode is a Japanese mission developed and launched by ISAS/JAXA, with NAOJ as domestic partner and NASA and STFC (UK) as international partners. It is operated by these agencies in co-operation with ESA and NSC (Norway).

\clearpage

\begin{deluxetable}{lccc}

\tablecaption{List of spectral lines used.\label{tab:eis_lines}}
\tablewidth{0pt}
\startdata
\tableline
\tableline

Ions &	Wavelength ($\mathring{\rm A}$) &	$\log (T [K])$\\
\tableline
\ion{Si}{7} & 272.65 & 5.70 \\
\ion{Si}{7} & 275.36 & 5.70\\
\ion{Si}{7} & 275.67 & 5.70 \\
\ion{Si}{7} (\textit{bl} \ion{Mg}{7}) & 278.45 & 5.70 \\
\ion{Si}{10} & 261.06 & 6.10 \\
\ion{Si}{10} & 277.26 & 6.10 \\
\tableline
\ion{Fe}{8} & 185.21 & 5.70  \\
\ion{Fe}{8} & 186.60 & 5.70  \\
\ion{Fe}{8} & 194.66 & 5.70  \\
\ion{Fe}{9} & 188.50 & 5.75 \\
\ion{Fe}{9} & 189.94 & 5.75\\
\ion{Fe}{9} & 197.86 & 5.75 \\
\ion{Fe}{10} & 184.54 & 6.00\\
\ion{Fe}{10} & 190.04 & 6.00\\
\ion{Fe}{11} & 182.17 & 6.00 \\
\ion{Fe}{11} & 188.22 & 6.00\\
\ion{Fe}{11} & 189.12 & 6.00\\
\ion{Fe}{12} & 192.40 & 6.10\\
\ion{Fe}{12} (\textit{sbl}) & 195.12 & 6.10\\
\ion{Fe}{12} & 196.54 & 6.10 \\
\ion{Fe}{13} & 202.04 & 6.15 \\
\ion{Fe}{15} & 284.16 & 6.30 \\
\tableline
\ion{S}{8} (\textit{bl} \ion{Fe}{11}) & 198.55 & 5.80\\

\enddata

\tablecomments{Emission lines are sorted by elements and ordered as a function of the peak temperature of their contribution functions. The blended and self-blended lines are indicated by the abbreviations \textit{bl} and \textit{sbl}, respectively.}
\end{deluxetable}

\clearpage

\begin{deluxetable}{lccccccccc}
\tablecaption{Identified plumes and interplumes; the location of their intensity maximum or minimum, respectively; and their
FIP-bias ratio, $f_{\rm (Fe\ \&\ Si)}/f_{\rm S}$, for the DEM analysis.\label{Tab:Results}}
\tablewidth{0pt} 
\startdata
\tableline
\tableline

 & \multicolumn{8}{c}{Data set} \\
\cline{2-9}
 & \multicolumn{2}{c}{1} & &\multicolumn{2}{c}{2} & &\multicolumn{2}{c}{3}\\
\cline{2-3} \cline{5-6} \cline{8-9}
Region & Location & $f_{\rm (Fe\ \&\ Si)}/f_{\rm S}$ & & Location & $f_{\rm (Fe\ \&\ Si)}/f_{\rm S}$ & & Location & $f_{\rm (Fe\ \&\ Si)}/f_{\rm S}$ &\\
\tableline

Plume & $\approx -80''$ & $\approx 0.73$ & &  & & & $\approx -100''$  & $\approx 0.78$ &\\

Interplume & $\approx -50''$ & $\approx 0.71$ & &  & & & $\approx -70''$  & $\approx 0.71$ &\\

Plume & $\approx -15''$ & $\approx 0.94$ & & $\approx -20''$ & $\approx 0.88$ & & $\approx -30''$  & $\approx 0.78$ &\\

Interplume &  &  & & $\approx 10''$ & $\approx 0.72$ & & $\approx 5''$ & $\approx 0.71$ &\\

Plume & $\approx 15''$ & $\approx 0.71$ & &  &  & &  &  &\\

Interplume & $\approx 60''$ & $\approx 0.69$ & &  & & & $\approx 50''$  & $\approx 0.70$ &\\

Plume & $\approx 85''$ & $\approx 0.80$ & &  & & & $\approx 100''$  & $\approx 0.79$ &\\

\tableline
\enddata

\end{deluxetable}

\clearpage

\begin{deluxetable}{lccccccccc}
\rotate
\tablecaption{Identified plumes and interplumes; the location of their intensity maximum or minimum, respectively; and their
FIP-bias ratio, $f_{\rm Si}/f_{\rm S}$, follow by the $1\sigma$ errors for the intensity, density and temperature, respectively,  for the isothermal analysis.\label{Tab:Results_iso}}
\tablewidth{0pt}
\startdata
\tableline
\tableline

 & \multicolumn{8}{c}{Data set} \\
\cline{2-9}
 & \multicolumn{2}{c}{1} & &\multicolumn{2}{c}{2} & &\multicolumn{2}{c}{3}\\
\cline{2-3} \cline{5-6} \cline{8-9}
Region & Location & $f_{\rm Si}/f_{\rm S}$ & & Location & $f_{\rm Si}/f_{\rm S}$ & & Location & $f_{\rm Si}/f_{\rm S}$ &\\
\tableline
Plume & $\approx -80''$ & $0.95 \pm 0.08, 0.01, 0.46$ & &  & & & $\approx -100''$  & $ 1.08 \pm 0.08, 0.01, 0.49$ &\\

Interplume & $\approx -50''$ & $1.05 \pm 0.10, 0.01, 0.54$ & &  & & & $\approx -70''$  & $0.96 \pm 0.11, 0.02, 0.52$ &\\

Plume & $\approx -15''$ & $ 1.32 \pm 0.10, 0.01, 0.60$ & & $\approx -20''$ & $ 1.23 \pm 0.08, 0.03, 0.53$ & &  $\approx -30''$ & $1.09 \pm 0.12, 0.02, 0.61$ &\\

Interplume &  &  & & $\approx 10''$ & $ 1.12 \pm 0.08, 0.02, 0.46$ & &$\approx 5''$ & $ 0.99 \pm 0.09, 0.01, 0.52$ &\\

Plume & $\approx 15''$ & $1.02 \pm 0.09, 0.01, 0.51$  & &  & & &  &  &\\

Interplume & $\approx 60''$ & $0.97 \pm 0.11, 0.01, 0.50$ & &  & & & $\approx 50''$  & $ 1.03 \pm 0.10, 0.02, 0.46$ &\\

Plume & $\approx 85''$ & $ 1.12 \pm 0.10, 0.02, 0.57$ & &  & & & $\approx 100''$  & $ 1.25 \pm 0.11, 0.02, 0.53$ &\\

\tableline
\enddata

\end{deluxetable}

\clearpage

\clearpage

\clearpage

\clearpage

\clearpage

\clearpage

\clearpage

\clearpage

\clearpage

\clearpage

\clearpage

\clearpage

\end{document}